\documentclass[twocolumn]{svjour3}          
\smartqed  
\usepackage{graphicx}
\usepackage{comment}
\usepackage{amsmath}
\usepackage{amssymb} 
\usepackage{color}
\usepackage{caption}
\usepackage{epsfig}
\usepackage{tabularx}
\usepackage{subfigure}
\usepackage{multicol}
\usepackage{multirow}
\usepackage{booktabs}
\usepackage{natbib}
\usepackage{arydshln}
\usepackage{tabulary}
\usepackage{floatrow}
\usepackage{url}
\usepackage{wrapfig}
\usepackage{arydshln}
\usepackage[misc]{ifsym}
\usepackage{etoolbox}
\newcommand*{\affaddr}[1]{#1} 
\newcommand*{\affmark}[1][*]{\textsuperscript{#1}}

\journalname{IJCV}
\begin{document}

\title{Image Quality Assessment for Perceptual Image Restoration: A New Dataset, Benchmark and Metric
\thanks{Dataset and codes are available at \url{https://www.jasongt.com/projectpages/pipal.html}.}
}

\author{%
    Jinjin Gu \affmark[1]\and
    Haoming Cai \affmark[2,3]\and
    Haoyu Chen \affmark[2]\and
    Xiaoxing Ye \affmark[2]\and\\
    Jimmy S. Ren \affmark[4,5]\and
    Chao Dong \affmark[3,6]
}
\authorrunning{Jinjin Gu \and Haoming Cai \and Haoyu Chen \and Xiaoxing Ye \and Jimmy S. Ren \and Chao Dong}

\institute{
    Jinjin Gu \at\email{jinjin.gu@sydney.edu.au}
    \and
    Haoming Cai \at\email{haomingcai@link.cuhk.edu.cn}
    \and
    Haoyu Chen \at\email{haoyuchen@link.cuhk.edu.cn}
    \and
    Xiaoxing Ye \at\email{xiaoxingye@link.cuhk.edu.cn}
    \and
    Jimmy S. Ren \at\email{rensijie@sensetime.com}
    \and
    Chao Dong \at\email{chao.dong@siat.ac.cn}
    \and
    \affaddr{\affmark[1] School of Electrical and Information Engineering, The University of Sydney.\\}
    \affaddr{\affmark[2] The Chinese University of Hong Kong, Shenzhen.\\}
    \affaddr{\affmark[3] ShenZhen Key Lab of Computer Vision and Pattern Recognition, SIAT-SenseTime Joint Lab, Shenzhen Institutes of Advanced Technology, Chinese Academy of Sciences.\\}
    \affaddr{\affmark[4] SenseTime Research.\\}
    \affaddr{\affmark[5] Qing Yuan Research Institute, Shanghai Jiao Tong University, Shanghai, China\\}
    \affaddr{\affmark[6] SIAT Branch, Shenzhen Institute of Artificial Intelligence and Robotics for Society.}
}

\date{}
\maketitle

\begin{abstract}
Image quality assessment (IQA) is the key factor for the fast development of image restoration (IR) algorithms.
The most recent perceptual IR algorithms based on generative adversarial networks (GANs) have brought in significant improvement on visual performance, but also pose great challenges for quantitative evaluation.
Notably, we observe an increasing inconsistency between perceptual quality and the evaluation results.
We present two questions:
(1) Can existing IQA methods objectively evaluate recent IR algorithms?
(2) With the focus on beating current benchmarks, are we getting better IR algorithms?
To answer the questions and promote the development of IQA methods, we contribute a large-scale IQA dataset, called Perceptual Image Processing ALgorithms (PIPAL) dataset. 
Especially, this dataset includes the results of GAN-based IR algorithms, which are missing in previous datasets.
We collect more than 1.13 million human judgements to assign subjective scores for PIPAL images using the more reliable ``Elo system''.
Based on PIPAL, we present new benchmarks for both IQA and super-resolution methods.
Our results indicate that existing IQA methods cannot fairly evaluate GAN-based IR algorithms.
While using appropriate evaluation methods is important, IQA methods should also be updated along with the development of IR algorithms.
At last, we shed light on how to improve the IQA performance on GAN-based distortion.
Inspired by the find that the existing IQA methods have an unsatisfactory performance on the GAN-based distortion partially because of their low tolerance to spatial misalignment, we propose to improve the performance of an IQA network on GAN-based distortion by explicitly considering this misalignment.
We propose the Space Warping Difference Network, which includes the novel $l_2$ pooling layers and Space Warping Difference layers.
Experiments demonstrate the effectiveness of the proposed method.
\keywords{Image Restoration \and Perceptual Image Restoration \and Image Quality Assessment \and Perceptual Super-Resolution \and Generative Adversarial Network}
\end{abstract}

\section{Introduction}
\label{sec:introduction}
Image restoration (IR) is a classic low-level vision problem, which aims at reconstructing high-quality images from distorted low-quality inputs.
Typical IR tasks include image super-resolution (SR), denoising, compression artifacts reduction, etc.
The whirlwind of progress in deep learning has produced a steady stream of promising IR algorithms that could generate less-distorted or perceptual-friendly images.
Nevertheless, one of the key bottlenecks that restrict the future development of IR methods is the ``evaluation mechanism''.
Although it is nearly effortless for human eyes to distinguish perceptually better images, it is challenging for an algorithm to fairly measure visual quality.
In this work, we will focus on the analysis of existing evaluation methods and propose a new image quality assessment (IQA) dataset, which not only includes the most recent IR methods but also has the largest scale/diversity.
The motivation will be first stated as follows.

IR methods are generally evaluated by measuring the similarity between the reconstructed images and ground-truth images via IQA metrics, such as PSNR \citep{psnr} and SSIM \citep{ssim}.
Recently, some non-reference IQA methods, such as Ma \citep{ma2017learning} and Perceptual Index (PI) \citep{blau2018perception}, are introduced to evaluate the recent perceptual-driven methods. 
To some extent, these IQA methods are the chief reason for the considerable progress of the IR field.
However, while new algorithms have been continuously improving IR performance, we notice an increasing inconsistency between quantitative results and perceptual quality.
For example, literature \citep{blau2018perception} reveals that the superiority of PSNR values does not always in accord with better visual quality. 
Although \citet{blau20182018} suggest that PI is more relevant to human judgement, algorithms with high PI scores (e.g., ESRGAN \citep{wang2018esrgan} and RankSRGAN \citep{zhang2019ranksrgan}) could still produce images with obvious unrealistic artifacts.
These conflicts lead us to rethink the evaluation methods for IR tasks.

An important reason for this situation is the invention of Generative Adversarial Networks (GANs) \citep{goodfellow2014generative} and GAN-based IR methods \citep{wang2018esrgan,gu2019image}, which brings completely new characteristics to the output images.
In general, these methods often fabricate seemingly realistic yet fake details and textures.
This presents a great challenge for existing IQA methods, which cannot distinguish the GAN-generated textures from noises and real details.
Then we naturally arise two questions:
(1) Can existing IQA methods objectively evaluate current IR methods, especially GAN-based methods?
(2) With the focus on beating benchmarks on the flawed IQA methods, are we getting better IR algorithms?
A few works have made early attempts to answer these questions by proposing new benchmarks for IR and IQA methods.
\citet{yang2014single} conduct a comprehensive evaluation of traditional SR algorithms.
\citet{blau2018perception} analyze the perception-distortion trade-off phenomenon and suggest the use of multiply IQA methods.
However, these prior studies usually apply unreliable human ratings of image quality and are generally insufficient in IR/IQA methods.
Especially, the results of GAN-based methods are missing in the above works.

To touch the heart of this problem, we need to have a better understanding of the new challenges brought by GAN.
The first issue is to build a new IQA dataset with the outputs of GAN-based algorithms.
An IQA dataset includes a lot of distorted images with visual quality levels annotated by humans.
It can be used to measure the consistency of the prediction of the IQA method and human judgement.
In this work, we contribute a novel IQA dataset, namely Perceptual Image Processing ALgorithms dataset (PIPAL).
The proposed PIPAL dataset distinguishes from previous datasets in four aspects:
(1) In addition to traditional distortion types (e.g., Gaussian noise/blur), PIPAL contains the outputs of several kinds of IR algorithms, including traditional algorithms, deep-learning-based algorithms, and GAN-based algorithms. In particular, this is the first time for the results of GAN-based algorithms to appear in an IQA dataset.
(2) We employ the Elo rating system \citep{elo1978rating} to assign subjective scores, which totally involves more than 1.13 million human judgements. Competing with existing rating systems (e.g., five gradations \citep{live} and Swiss system \citep{tid2008}), the Elo rating system provides much more reliable probability-based rating results. Furthermore, it is easily extensible, which allows users to update the dataset by directly adding new distortion types.
(3) The proposed dataset contains 29K images in total, including 250 high-quality reference images, and each of which has 116 distortions. To date, PIPAL is the largest IQA dataset with complete subjective scoring.
(4) We also contribute a new open-source web-based rating software called Image Quality Opinion Scoring (IQOS) system that allows users to assign subjective scores for their own images easily.

\begin{table*}[t!]
    \centering
    \caption{Comparison with the previous IQA datasets. A primary peculiarity that distinguishes the proposed PIPAL dataset and previous work is the including of the outputs of GAN-based algorithms. Although BAPPS dataset \citep{zhang2018unreasonable} contains much more reference patches (187.7k), each of them has only two distorted images, which greatly limits its reliability and accuracy on evaluating IR algorithms. The PieAPP dataset \citep{prashnani2018pieapp} records the probability of human preference between two images. However, such pairwise preference can not be summarized into a mean opinion score (MOS). Pairwise preference also leads to the requirement of a large number of human judgements. The proposed PIPAL dataset is the largest IQA dataset with MOS.}
    \label{tab:datasets}
    \resizebox{1.0\linewidth}{!}{
    \begin{tabular}{cccccccc}
        \multirow{2}{*}{\textbf{Dataset}}  &
        \# Ref. &
        Image &
        Distortion &
        \# Distort. &
        \# Distort. &
        \# Human &
        judgement \\
        & 
        images &
        types &
        types &
        types &
        images &
        judgements &
        type \\
        \hline
        LIVE &
        29 &
        image &
        traditional &
        5 &
        0.8k &
        25k &
        MOS (Five gradations) \\
        CSIQ &
        30 &
        image &
        traditional &
        6 &
        0.8k &
        5k &
        MOS (Direct ranking) \\
        TID2008 &
        25 &
        image &
        traditional &
        17 &
        1.7k &
        256k &
        MOS (Swiss system)\\
        TID2013 &
        25 &
        image &
        traditional &
        25 &
        3.0k &
        524k &
        MOS (Swiss system) \\
        BAPPS &
        187.7k &
        patch (64$\times$64) &
        trad. $+$ alg. outputs&
        425 &
        375.4k &
        484.3k &
        2 Alternative Choice \\
        PieAPP &
        200 &
        patch (256$\times$256) &
        trad. $+$ alg. outputs&
        75 &
        20.3k &
        2492k &
        Prob. of Preference \\
        \hline
        PIPAL &
        \multirow{2}{*}{250} &
        patch &
        trad. $+$ alg. outputs&
        \multirow{2}{*}{40} &
        \multirow{2}{*}{29k} &
        \multirow{2}{*}{1.13m} &
        MOS \\
        (Ours)&
        &
        (288$\times$288)&
        \emph{including GAN} &
        &
        &
        &
        (Elo rating system)\\
        \hline
    \end{tabular}
    }
\end{table*}{}  

With the PIPAL dataset, we are able to answer the aforementioned two questions.
First, we build a benchmark using the proposed PIPAL dataset for the existing IQA methods to explore whether can they objectively evaluate current IR methods.
Experiments indicate that PIPAL poses challenges for these IQA methods.
Evaluating IR algorithms only using existing metrics is not appropriate.
Our research also shows that compared with the widely-used metrics (e.g., PSNR and PI), PieAPP \citep{prashnani2018pieapp}, LPIPS \citep{zhang2018unreasonable} and WaDIQaM \citep{wadiqam} are relatively more suitable for evaluating IR algorithms, especially GAN-based algorithms.
We also study the characteristics and difficulties of GAN-based distortion by comparing them with some well-studied traditional distortions.
Based on the results, we argue that existing IQA methods' low tolerance toward spatial misalignment may be a key reason for their performance drop.
To answer the second question, we review the development of SR algorithms in recent years.
We find that none of the existing IQA methods is always effective in evaluating SR algorithms.
With the invention of new IR technologies, the corresponding evaluation methods also need to be adjusted to continuously promote the development of the IR field.

At last, we shed light on how to improve the IQA performance on GAN-based distortion.
We argue that the existing IQA methods have an unsatisfactory performance on the GAN-based distortion partially because of their low tolerance to spatial misalignment.
Inspired by this finding, we improve the performance of an IQA network on GAN-based distortion by explicitly considering this misalignment.
Firstly, we introduce $l_2$ pooling layers to replace the original pooling layers in IQA networks, which can avoid aliasing during downsampling and make the extracted image features shift-invariant.
Secondly, we propose a new feature comparison operation called Space Warping Difference (SWD) layer, which compares the features that not only on the corresponding position but also on a small range around it.
This operation explicitly makes the comparison robust to small displacements.
By employing the $l_2$ pooling layers and the SWD layers, we propose the space warping difference IQA network (SWDN).
Extensive experimental results demonstrate that the proposed components are effective and SWDN achieves state-of-the-art performance on GAN-based distortion.

We note that a shorter conference version of this paper appeared in \citet{pipal}.
In addition to the conference version, this manuscript includes the following additional contents:
(1) We include more details of the PIPAL dataset in Section~\ref{sec:pipal}, including examples of the reference images in \figurename~\ref{fig:references}, the distribution of the subjective scores in \figurename~\ref{fig:histo}, and a new image quality opinion scoring system with details about the rating process in Section~\ref{sec:sys}.
(2) We include more results and discussion in Section~\ref{sec:iqa:benchmarks} and \ref{sec:ir:benchmarks}. Especially, we present more details about the ``counter-example'' experiment in Section~\ref{sec:ir:benchmarks}.
(3) We introduce a new IQA network called SWDN in Section~\ref{sec:swdn}. In Section~\ref{sec:iqanet:exp}, we conduct both comparison and ablation experiments to demonstrate the effectiveness of the proposed layers and the SWDN network.

\section{Related Work}
\label{sec:related}

\paragraph{Image Restoration (IR).}
As a fundamental computer vision problem, IR aims at recovering a high-quality image from its degraded observations.
Some common degradation processes include nosing, blurring, downsampling, etc.
For different types of image degradation, there are corresponding IR tasks, respectively.
For example, image SR aims at recovering the high-resolution image from its low-resolution observation and image denoising aims at removing unpleasant noise.
In past decades, plenty of IR algorithms have been proposed to continuously improve the performance.
The early algorithms use hand-craft features \citep{bm3d,ywhm2010} or exploit image priors \citep{tsg2013,a+2014} in optimization problems to reconstruct images.
Since the pioneer work of using Convolution Neural Networks (CNNs) to learn the IR mapping \citep{jain2009natural,srcnn2014}, the deep-learning-based algorithms have dominated IR research due to their remarkable performance and usability.
Recently, with the invention of Generative Adversarial Networks (GANs) \citep{goodfellow2014generative}, GAN-based IR methods \citep{enhancenet2017,zhang2019ranksrgan,gu2019image} are not limited to getting a higher PSNR performance but trying to have better perceptual effect.
However, these image restoration algorithms are not perfect.
The results of those algorithms also include various image defects, and they are different from the traditional distortions that are often discussed in previous IQA researches.
With the development of IR algorithms and the emergence of new technologies (e.g., GAN-based algorithms), evaluating the results of these algorithms becomes more and more challenging.
In this paper, we mainly focus on the restoration of low-resolution images, noisy images, and images degraded by both resolution reduction and noise.

\paragraph{Image Quality Assessment (IQA).}
The IQA methods were developed to measure the perceptual quality of images that may be degraded during acquisition, compression, reproduction, and post-processing operations.
According to different usage scenarios, IQA methods can be divided into full-reference methods (FR-IQA) and no-reference methods (NR-IQA).
FR-IQA methods measure the similarity between two images from the perspective of information or perceptual feature similarity, and have been widely used in the evaluation of image/video coding, restoration and communication quality.
Beyond the most widely-used PSNR, FR-IQA methods follow a long line of works that can trace back to SSIM \citep{ssim}, which first introduces structural information in measuring image similarity.
After that, various FR-IQA methods have been proposed to bridge the gap between results of IQA methods and human judgements \citep{fsim,ifc,vsi}.
Similar to other computer vision problems, advanced data-driven methods have also motivated the investigation of applications of IQA \citep{zhang2018unreasonable,prashnani2018pieapp}.
In addition to the above FR-IQA methods, NR-IQA methods are proposed to assess image quality without a reference image.
Some popular NR-IQA methods include NIQE \citep{niqe}, \citet{ma2017learning}, BRISQUE \citep{brisque}, and PI \citep{blau2018perception}.
In some recent works, NR-IQA and FR-IQA methods are combined to measure IR algorithms \citep{blau2018perception}.
Despite of the the progress of IQA methods, only few IQA methods (e.g., PSNR, SSIM and PI) are used to evaluate IR methods.

\paragraph{IQA Datasets.}
In order to evaluate and develop IQA methods, many datasets have been proposed, such as LIVE \citep{live}, CSIQ \citep{csiq}, TID2008 and TID2013 \citep{tid2008,tid2013}.
These datasets provide both distorted images and the corresponding subjective scores, and they have served as baselines for evaluation of IQA methods.
These datasets are mainly distinguished from each other in three aspects: (1) the collecting of the reference images, (2) the number of distortions included and their types and (3) the collection strategy of subjective score.
A quick comparison of these datasets can be found in \tablename~\ref{tab:datasets}.
In addition, there are also some perceptual similarity datasets such as PieAPP \citep{prashnani2018pieapp}, and BAPPS \citep{zhang2018unreasonable}, which only contain pairwise judgements of distorted images.

\begin{table}[t!]
    \centering
    \footnotesize
    \caption{The distortion types in our dataset. The traditional distortions are performed by basic low-level image editing operations.
    For the algorithms distortions, we select classic algorithms and representative new algorithms in recent years. For the algorithms with visually similar outputs (it happens in algorithms where the PSNR values are very close), we only select one for each year as a representative.
    \emph{In particular, we include 19 different GAN-based algorithms distortions.}}
    \label{tab:distortions}
    \resizebox{1.0\linewidth}{!}{
    \begin{tabular}{p{0.25\linewidth}<{\centering}p{0.75\linewidth}<{\centering}}
        \textbf{Sub-type} & \textbf{Distortion Types} \\
        \hline
        Traditional & Gaussian blur, motion blur, image compression, Gaussian noise, spatial warping, bilateral filter, comfort noise.\\
        \hline
        Super-Resolution & interpolation method, traditional methods, SR with kernel mismatch, PSNR-oriented methods, GAN-based methods. \\
        \hline
        Denoising & mean filtering, traditional methods, deep learning-based methods. \\
        \hline
        Mixture Restoration & SR of noisy images, SR after denoising, SR after compression noise removal \\
        \hline
    \end{tabular}
    }
\end{table}{}  

\section{Perceptual Image Processing ALgorithms (PIPAL) Dataset}
\label{sec:pipal}

We first describe the peculiarities of the proposed dataset from the aforementioned aspects of (1) the collecting of the reference images, (2) the number of distortions and their types, and (3) the collecting of subjective score, respectively.
Then, we present a new image quality opinion scoring (IQOS) system for collecting user ratings.

\begin{figure}[t]
    \centering
    \includegraphics[width=1.0\linewidth]{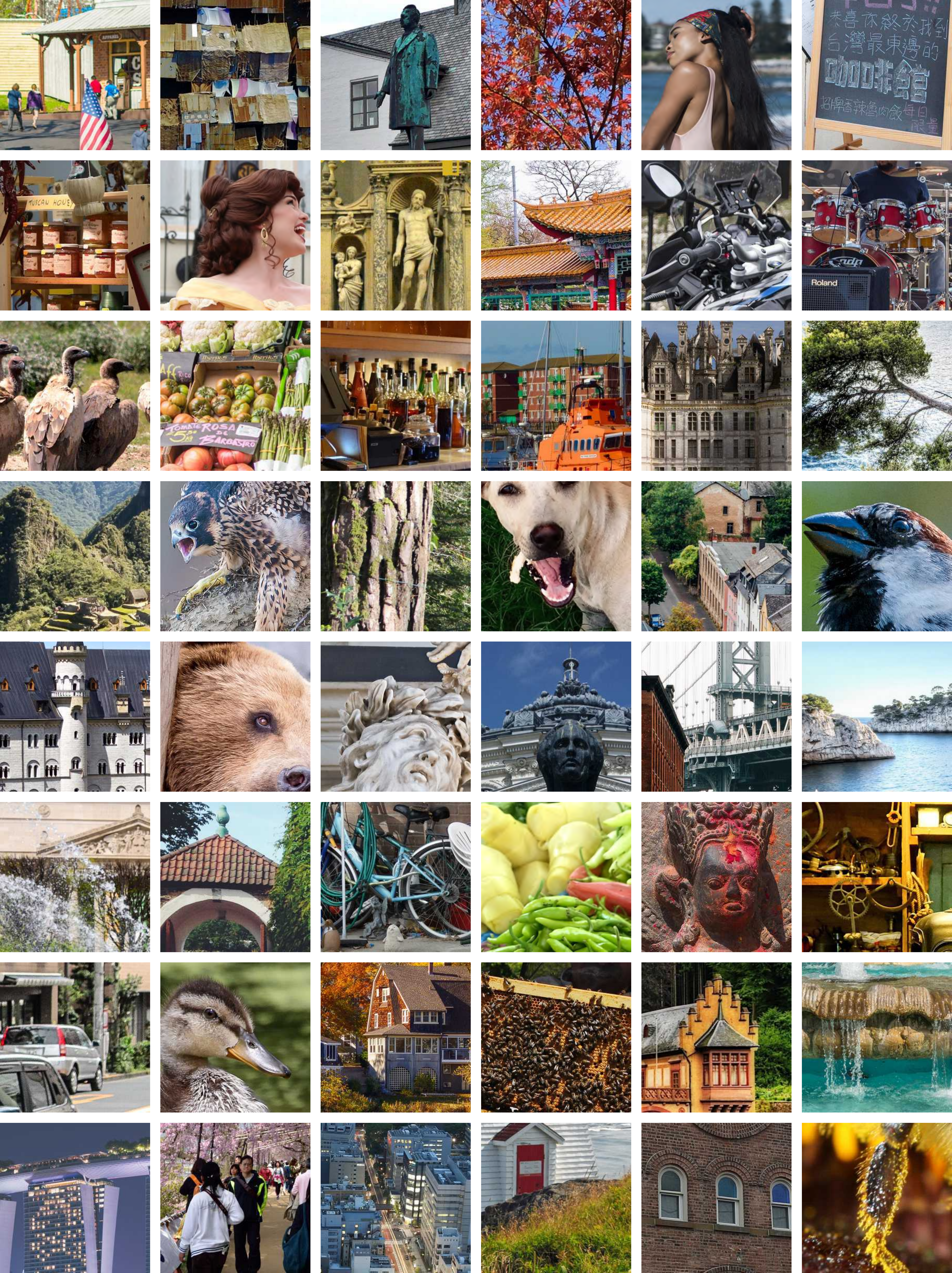}
    \caption{Examples of the reference images. We include a wide variety of textures, such as buildings, trees, grasses, animal fur, human faces, text, and artificial textures.}
    \label{fig:references}
\end{figure}  

\begin{figure*}[t]
    \centering
    \includegraphics[width=1.0\linewidth]{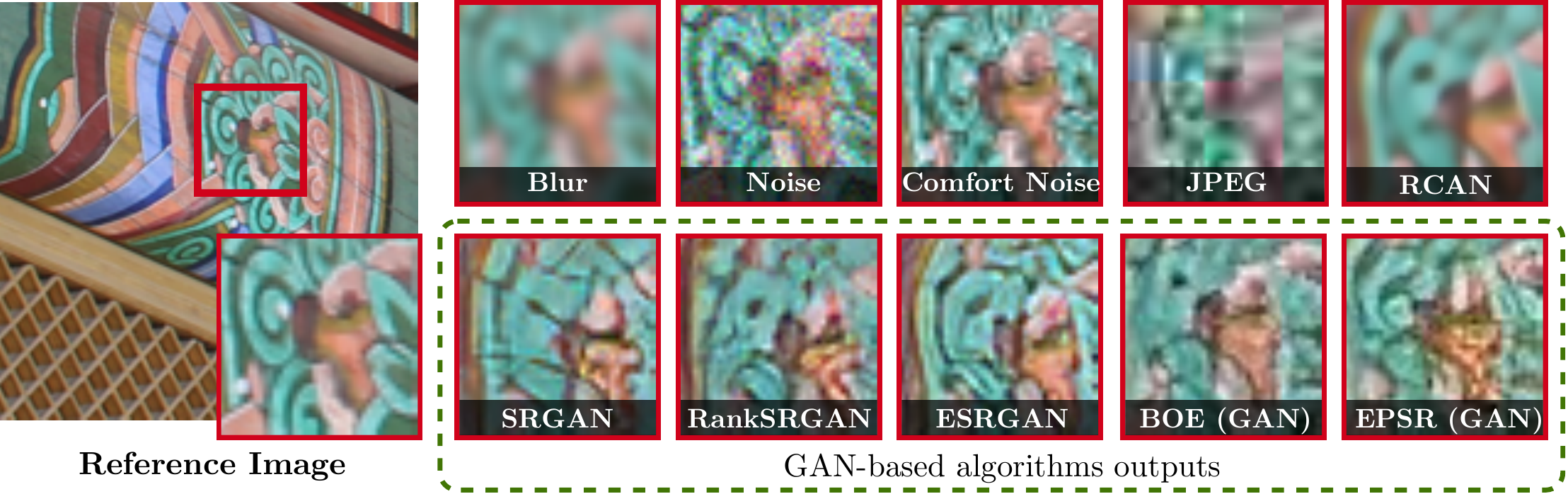}
    \caption{GAN-based SR distortions. Unlike the distortions in the upper row, which do not follow the natural image distribution. One can see that the GAN-based distortions are similar to natural images. However, their details are wrong and different from the reference image.}
    \label{fig:GANvis}
\end{figure*}  

\subsection{Collection of reference images.}
We select 250 image patches from two high-quality image datasets (DIV2K \citep{div2k} and Flickr2K \citep{timofte2017ntire}) as the reference images.
When selecting them, we mainly focus on the representative texture areas that are relatively hard to restore.
\figurename~\ref{fig:references} shows an overview of the selected reference images.
As one can see, the selected reference images are representative of a wide variety of real-world textures, including but not limited to buildings, trees, grasses, animal fur, human faces, text, and artificial textures.
In our dataset, the images are of size $288\times288$, which could meet the requirements of most IQA methods.

\subsection{Distortion Types.}
In our dataset, we have 40 distortion types, which can be divided into four sub-types.
An over-view of these distortion types is shown in \tablename~\ref{tab:distortions}.
The first sub-type includes some traditional distortions (e.g., blur, noise, and compression), which are usually performed by basic low-level image editing operations.
In the previous datasets, these distortions can be very severe, such as PieAPP and TID2013.
However, in our dataset, we constrain the situation of severe distortions as we want these distortions to be comparable to the IR results, which are not likely to be very low-quality.
The second sub-type includes the SR results from existing algorithms.
Although some recent datasets \citep{zhang2018unreasonable,prashnani2018pieapp} have covered some of the SR results, they contain results that are inferior in algorithms number and types to our dataset.
We divide the selected SR algorithms into three categories -- traditional algorithms, PSNR-oriented algorithms, and GAN-based algorithms.
The results of traditional algorithms can be understood, to some extent, as loss of detail.
The PSNR-oriented algorithms are usually based on deep-learning technology.
Comparing with the traditional algorithms, their outputs tend to have sharper edges and higher PSNR performance.
The outputs of GAN-based algorithms are more complicated and challenging for IQA methods.
They do not quite match the quality of detail loss, as they usually contain texture-like noises, or the quality of noise, the texture-like noise is similar to the ground truth in appearance but is not accurate.
An example of GAN-based distortions is shown in \figurename~\ref{fig:GANvis}.
Measuring the similarity of incorrect yet similar features is of great importance to the development of perceptual SR.
The third sub-type includes the outputs of several denoising algorithms.
Similar to image SR, the used denoising algorithms contain both model-based algorithms and deep-learning-based algorithms.
In addition to Gaussian noise, we also include JPEG compression noise removal results.
At last, we include the restoration results of the mixed degradation.
As revealed in \citep{zhang2018learning,qian2019trinity}, performing denoising and SR sequentially will bring new artifacts or different blur effects that barely occur in other IR tasks.

In summary, we have 40 different distortion types and 116 different distortion levels, resulting in 29k distorted images in total.
Note that although the number of distortion types is less than some of the existing datasets, we contain a lot of new distortion types and, especially, a large number of IR algorithms' results and GAN results.
This allows the proposed dataset to provide a more objective benchmark for not only IQA methods but also IR methods.

\subsection{Elo Rating for Mean Opinion Score.}
Given distorted images, the Mean Opinion Score (MOS) is provided for each distorted image.
In literature, there are several methodologies used to assess the visual quality of an image \citep{tid2013,zhang2018unreasonable}.
Early datasets \citep{sheikh2006statistical} use five gradations rating method.
Images are assigned into five categories such as ``Bad'', ``Poor'', ``Fair'', ``Good'', and ``Excellent''.
This method may lead to a huge bias when the raters do not have enough experience.
In recent years, datasets usually collect MOS through a large number of pairwise selections using the Swiss rating system \citep{tid2013}.
However, the way this pairwise MOS is calculated makes it dependent on a specific dataset, which means that the MOS scores of two distorted images can change significantly when they are included in two different datasets.
In order to eliminate this set-dependence effect, \citet{prashnani2018pieapp} propose to build a dataset only based on the probability of pairwise preference, which provides a more accurate propensity probability.
However, it not only requires a large number of human judgements, but also can not provide the MOS scores for distortion types, which are important for building benchmark for IR algorithms.
In the proposed dataset, we employ the Elo rating system \citep{elo1978rating} to bring pairwise preference probability and rating system together.
The use of the Elo system provides reliable human ratings with fewer human judgements.

The Elo rating system is a statistic-based rating method and was first proposed for assessing chess player levels.
We assume the rater preference between two images $I_A$ and $I_B$ follows a Logistic distribution parameterized by their Elo Scores \citep{elo20088}.
Given their Elo scores $R_A$ and $R_B$, the expected probability that one user would prefer $I_A$ to $I_B$ is given by:
\begin{equation}
    P_{A>B}=\frac{1}{1+10^{(R_B-R_A)/M}},
    \label{eq:elo}
\end{equation}
where $M$ is the parameter of the distribution.
In our dataset we use $M=400$, which is a widely used setting in chess games.
Then, the probability that one user would prefer $I_B$ to $I_A$ is in a symmetrical form:
\begin{equation}
    P_{B>A}=\frac{1}{1+10^{(R_A-R_B)/M}}.
\end{equation}
Obviously, we have $P_{A>B}+P_{B>A}=1$.
Once a rater makes a choice, we then update the Elo score for both $I_A$ and $I_B$ use the following rules:
\begin{align}
    &R_A'=R_A+K\times(S_A-P_{A>B}),\\&R_B'=R_B+K\times(S_B-P_{B>A}),
\end{align}
where $K$ is the max value of the Elo score change in one step.
In our dataset, $K$ is set to 16.
$S_A$ indicates whether $I_A$ is chosen: $S_A=1$ if $I_A$ wins and $S_A=0$ if $I_A$ fails.
With thousands of human judgements, the Elo scores for each distorted images will converge.
The average of the Elo scores in the last few steps will be assigned as the MOS subjective score.
The averaging operation aims at reducing the randomness of the Elo score changes.

\begin{figure}[t]
    \centering
    \includegraphics[width=1.0\linewidth]{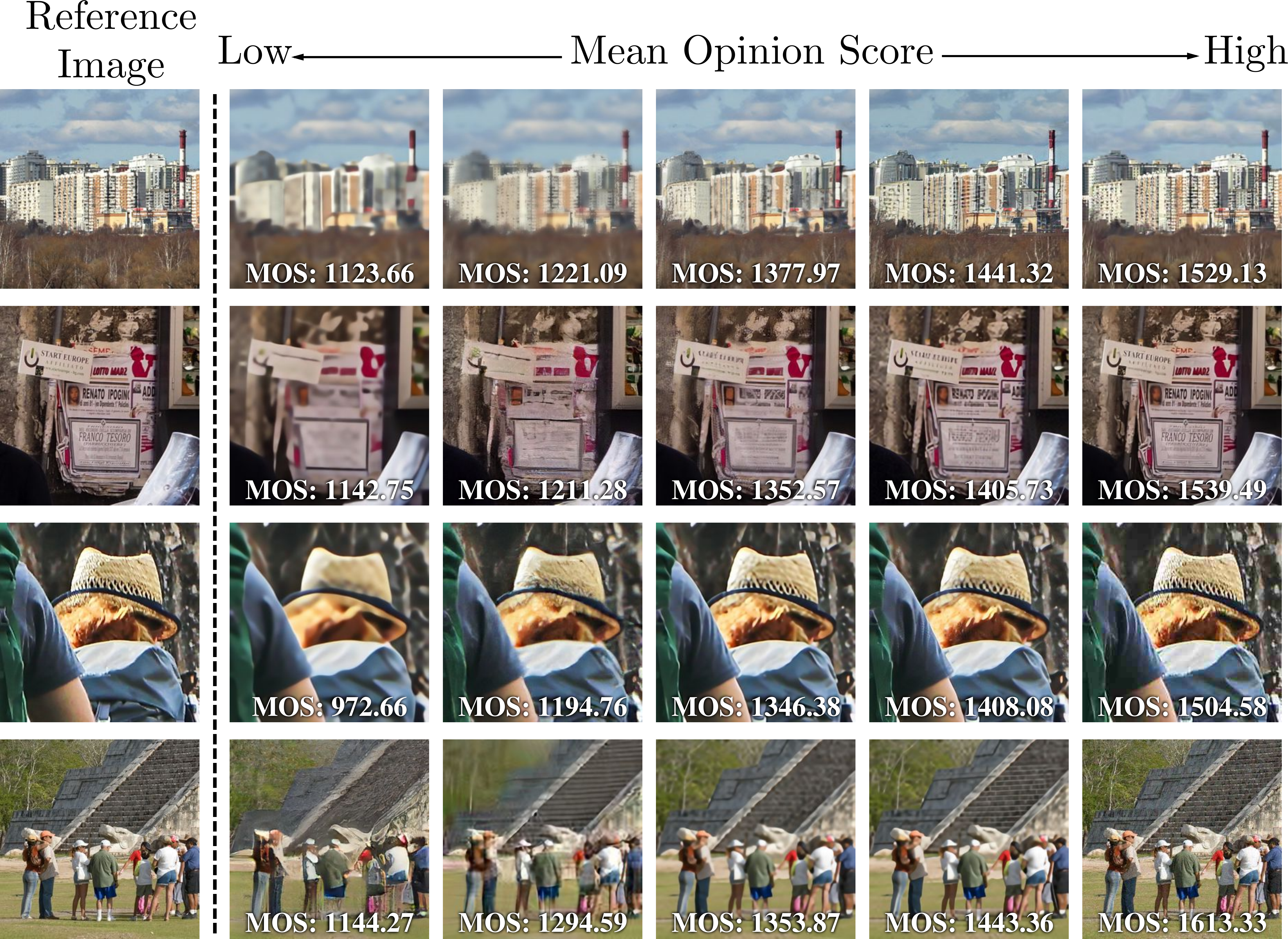}
    \caption{Examples of distorted images and their MOS values. The MOS value and the perceptual similarity are positively related.}
    \label{fig:mosvis}
\end{figure}

\begin{figure}[t]
    \centering
    \includegraphics[width=1.0\linewidth]{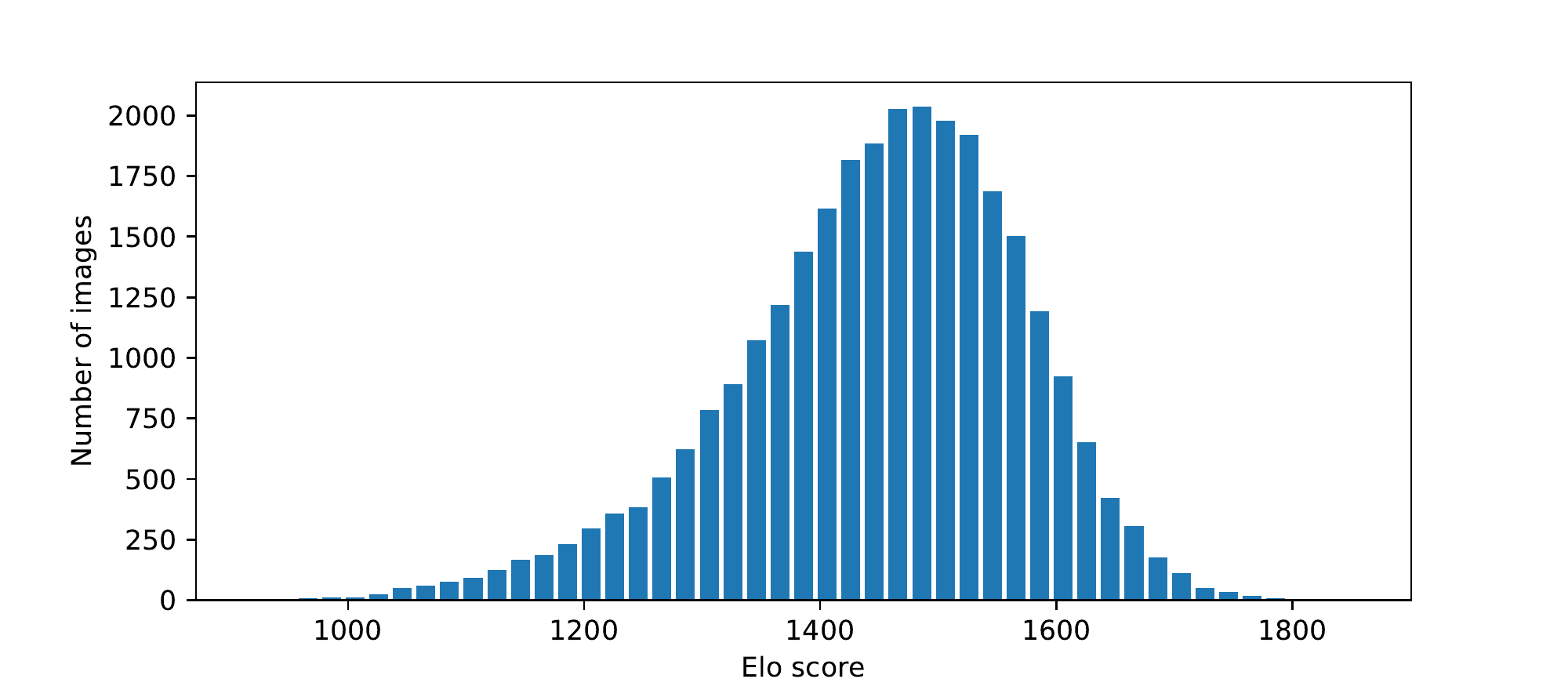}
    \caption{The MOS histogram for PIPAL dataset. The MOS values are mainly concentrated in the interval from 1300 to 1600.}
    \label{fig:histo}
\end{figure}

An example might help understand the Elo system.
Suppose that $R_A=1500$ and $R_B=1600$, then we have $P_{A>B}\approx0.36$ and $P_{B>A}\approx0.64$.
In this situation, if $I_A$ is chosen, the updated Elo score for $I_A$ will be $R_A'=1500 + 16\times(1-0.36)\approx1510$ and the new score for $I_B$ is $R_B'=1600 + 16\times(0-0.64)\approx1594$; if $I_B$ is chosen, the new score will be $R_A'\approx1494$ and $R_B'\approx1605$.
Note that because the expected probabilities for different images being chosen are different, the value changes of the Elo scores are different.
This indicates that when the quality is too different, the winner will not get a lot from winning the bad image.
According to Eq.\eqref{eq:elo}, a score difference of 200 indicates 76\% chance to win, and 400 indicates the chance more than 90\%.
At first, we assign an Elo score of 1400 for each distorted image.
After numerous human judgements (in our dataset, we have 1.13 million human judgements), the Elo score for each image are collected.
\figurename~\ref{fig:mosvis} shows some examples of distortions and their corresponding MOS scores, and \figurename~\ref{fig:histo} shows the distribution of MOS scores in PIPAL dataset.

Another superiority of employing the Elo system is that our dataset could be dynamic and can be extend in the future.
The Elo system has been widely used to evaluate the relative level of players in electronic games \citep{wang2015thinking}, where the players are constantly changing and the Elo system can provide ratings for new players in a few gameplays.
Recall that one of the chief reason that these IQA methods are facing challenges is the invention of GAN and GAN-based IR methods.
What if other novel image generation technologies are proposed in the future?
Do people need to build a new dataset to include those new algorithms?
With the extendable characteristic of the Elo system, one can easily add new distortions into this dataset and follow the same rating process.
The Elo system will automatically adjust the Elo score for all the distortions without re-rating for the old ones.

\begin{table*}[t]
    \centering
    \footnotesize
    \resizebox{1.0\linewidth}{!}{
    \begin{tabular}{
    p{2.0cm}
    p{2.1cm}<{\centering}
    p{2.1cm}<{\centering}
    p{2.1cm}<{\centering}
    p{2.1cm}<{\centering}
    p{2.1cm}<{\centering}
    p{2.1cm}<{\centering}}
    \toprule
        \multirow{2}{*}{\textbf{Method}} &
        \textbf{Traditional} &
        \multirow{2}{*}{\textbf{Denoising}} &
        \multirow{2}{*}{\textbf{SR Full}} &
        \textbf{Traditional} &
        \textbf{PSNR.} &
        \textbf{\textit{GAN-based}}
        \\
         &
        \textbf{Distortion} &
         &
         &
        \textbf{SR} &
        \textbf{SR} &
        \textbf{\textit{SR}}
        \\
        \hline
        PSNR $\uparrow$         &	0.3589 	&	0.4542 	&	0.4099 	&	0.4782 	&   0.5462 	&	0.2839  \\
        NQM $\uparrow$          &	0.2561 	&	0.5650 	&	0.4742 	&	0.5374 	&	0.6462 	&	0.3410  \\
        UQI $\uparrow$          &	0.3455 	&	0.6246 	&	0.5257 	&	0.6087	&	0.7060 	&	0.3385  \\
        SSIM $\uparrow$         &	0.3910 	&	0.6684 	&	0.5209 	&	0.5856 	&	0.6897 	&	0.3388  \\
        MS-SSIM $\uparrow$      &	0.3967 	&	0.6942 	&	0.5596 	&	0.6527 	&   0.7528 	&	0.3823  \\
        IFC $\uparrow$          &	0.3708 	&	\textbf{0.7440} 	&	0.5651 	&	\textbf{0.7062} 	&	\textbf{0.8244} 	&	0.3217  \\
        VIF $\uparrow$          &	0.4516  &	\textbf{0.7282}     &	0.5917  &	\textbf{0.6927}  &	\textbf{0.7864}  &	0.3857  \\
        VSNR-FR $\uparrow$      &	0.4030  &	0.5938  &	0.5086  &	0.6146  &	0.7076  &	0.3128  \\
        RFSIM $\uparrow$        &	0.3450  &	0.4520  &	0.4232  &	0.4593  &	0.5525  &	0.2951  \\
        GSM $\uparrow$                             &	0.5645  &	0.6076  &	0.5361  &	0.6074  &	0.6904  &	0.3523  \\
        SR-SIM $\uparrow$                          &	\textbf{0.6036}  &	0.6727  &	0.6094  &	0.6561  &	0.7476  &	0.4631  \\
        FSIM $\uparrow$                            &	0.5760  &	0.6882  &	0.5896  &	0.6515  &	0.7381  &	0.4090  \\
        FSIM$_\mathcal{C}$ $\uparrow$              &	0.5724  &	0.6866  &	0.5872  &	0.6509  &	0.7374  &	0.4058  \\
        VSI $\uparrow$                             &	0.4993  &	0.5745  &	0.5475  &	0.6086  &	0.6938  &	0.3706  \\
        MAD $\downarrow$                           &	0.3769  &	0.7005  &	0.5424  &	0.6720  &	0.7575  &	0.3494  \\
        LPIPS-Alex $\downarrow$                    &	0.5935  &	0.6688  &	0.5614  &	0.5487  &	0.6782  &	\textbf{0.4882}  \\
        LPIPS-VGG $\downarrow$                     &	0.4087  &	0.7197  &	\textbf{0.6119}  &	0.6077  &	0.7329  &	0.4816  \\
        PieAPP $\downarrow$                        &	\textbf{0.6893}  &	\textbf{0.7435}  &	\textbf{0.7172}  &	\textbf{0.7352}  &	\textbf{0.8097}  &	\textbf{0.5530}  \\
        DISTS $\downarrow$                         &	\textbf{0.6213}  &	0.7190  &	\textbf{0.6544}  &	0.6685  &	0.7733  &	\textbf{0.5527}  \\
        \hdashline
        \underline{NIQE} $\downarrow$              &	0.1107  &	-0.0059  &	0.0320  &	0.0599  &	0.1521  &	0.0155  \\
        \underline{Ma \textit{et al.}} $\uparrow$                 &	0.4526  &	0.4963  &	0.3676  &	0.6176  &	0.7124  &	0.0545  \\
        \underline{PI} $\downarrow$                &	0.3631  &	0.3107  &	0.1953  &	0.4833  &	0.5710  &	0.0187  \\
    \toprule
    \end{tabular}}
    \caption{The SRCC results with respect to different distortion sub-types.
    $\uparrow$ means the higher the better while  $\downarrow$ means the lower the better. Higher coefficient matches perceptual score better. We indicate the top 3 performance with \textbf{blod} values.
    }
    \label{tab:iqa_results}
\end{table*}  

\subsection{Image Quality Opinion Scoring (IQOS) System.}
\label{sec:sys}
\begin{figure}[t]
    \centering
    \includegraphics[width=1.0\linewidth]{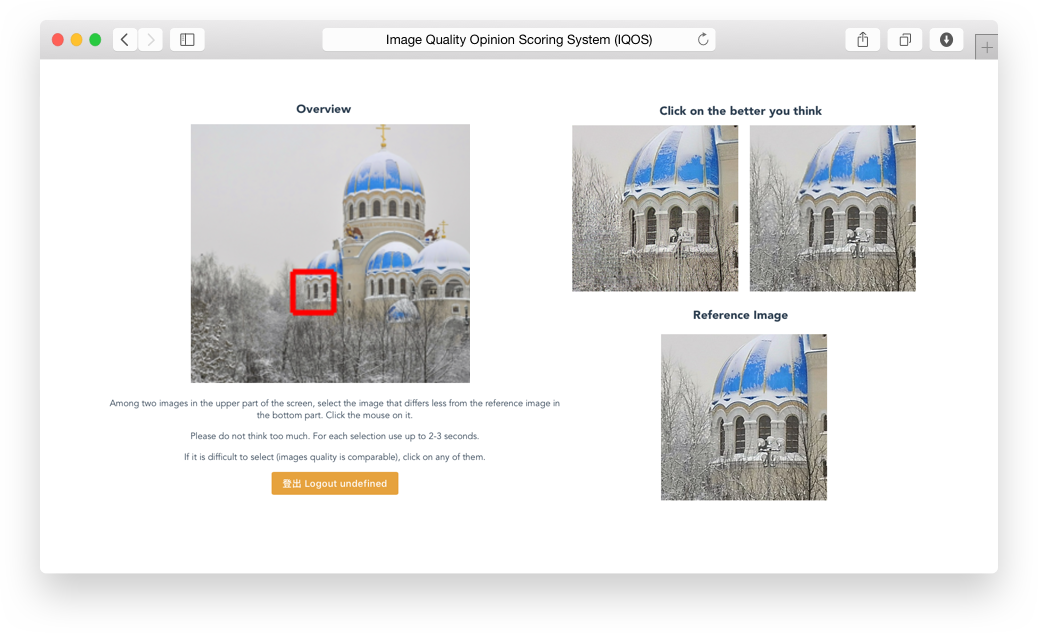}
    \caption{Screenshot of the user interface of the proposed image quality opinion scoring system.}
    \label{fig:iqos}
\end{figure}

Many prior works have proposed platform and software to collect subjective evaluations such as Amazon Mechanical Turk (MTurk) and Versus \citep{vuong2018versus}.
However, the Elo rating process is dynamic and can not be simply done with MTurk.
In this work, we develop a new web-based image quality opinion scoring system (IQOS), which integrates both Swiss rating system and Elo rating system.
The back-end of this system is powered by the \texttt{Flask} framework and the web-based user interface is powered by \texttt{Vue.js}.
A screenshot of the interface is shown in \figurename~\ref{fig:iqos}.
For each reference image, we show two of its distorted images.
The raters are required to choose the distorted image that differs less from the reference image by directly clicking on it.
We ask the raters to make decision in a short time.
For two images that can not be easily distinguished, we ask raters to choose any one of them randomly.
When the number of judgements is big enough, the probability of two images to be selected will be close.
The Elo system will assign images with very close Elo scores, this is consistent with the fact that they do have similar perceptual qualities.
We also provide an overview for each reference image to show its role in the original whole image.
This software is easy to use and we will make it open-source to contribute the community.

In our work, more than 500 raters have participated in the experiments both in the controlled laboratory environment and via Internet.
The raters are composed of both professional data annotation teams and volunteers.
The observation conditions are controlled within a reasonable range.
Note that we do not strictly follow the recommendation settings of ITU \citep{itu20121401}.
As these IQA methods are designed for measuring visual quality under unknown conditions in practice, visualizing and analyzing image quality under slightly changing conditions can provide a reasonably good verification of the IQA methods.
The non-identical conditions of the experiments take into account how to evaluate visual quality in real practice for different users.

\section{Results}
In this section, we conduct a comprehensive study using the proposed PIPAL dataset.
We first build a benchmark for IQA methods.
Through this benchmark, we can answer the question that ``can existing IQA methods objectively evaluate recent IR algorithms?''
We then build a benchmark for some recent SR algorithms to explore the relationship between the development of IQA methods and IR research.
We can get the answer of ``are we getting better IR algorithms by beating benchmarks on these IQA methods?''
At last, we study the characteristics of GAN-based distortion by comparing them with other existing distortion types.

\begin{figure*}
    \centering
    \includegraphics[width=0.9\linewidth]{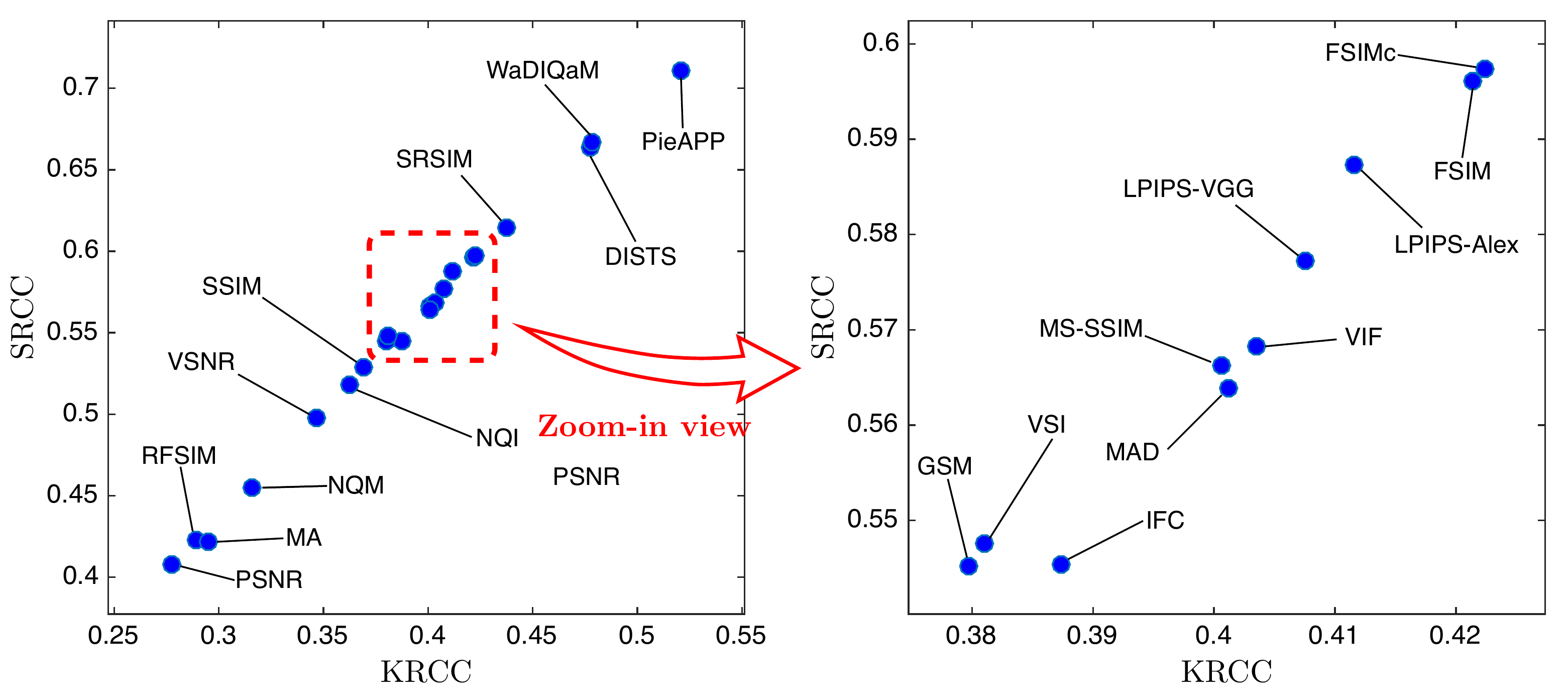}
  \caption{Quantitative comparison of IQA methods. The right figure is the zoom-in view. Higher coefficient matches perceptual score better.}
  \label{fig:rcc}
\end{figure*}

\begin{figure*}[t]
    \begin{center}
        \includegraphics[width=1.0\linewidth]{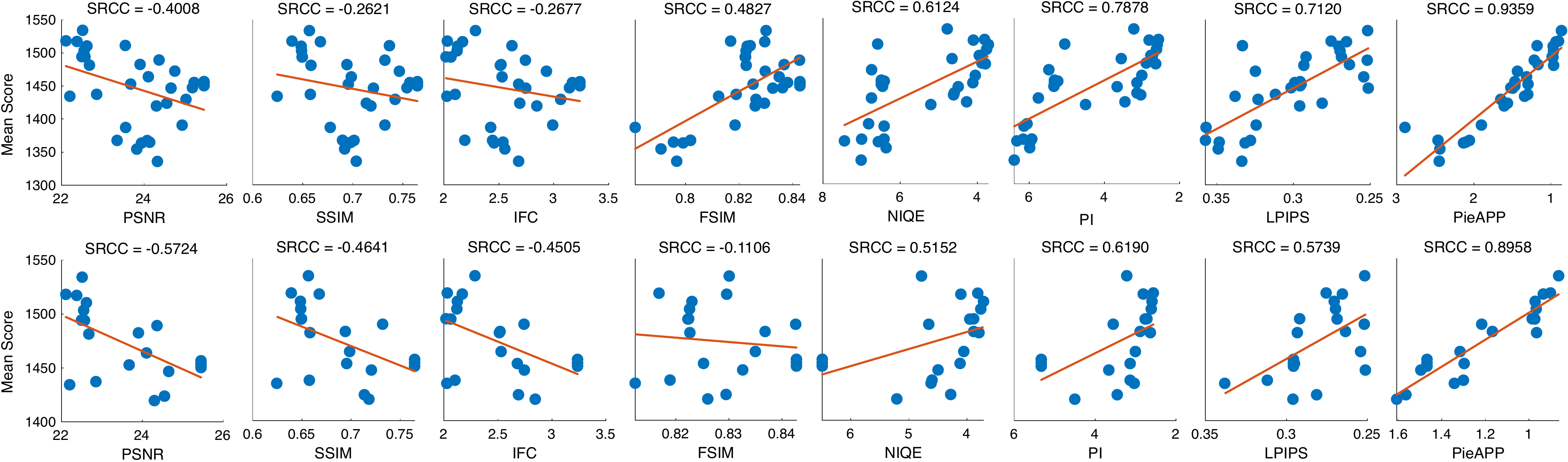}
    \caption{Analysis of IQA methods in evaluating IR methods. On the first row are the scatter plots of MOS score vs. IQA methods for all SR algorithms.
    On the second row are scatter plots for GAN-based SR algorithms.}
    \label{fig:sr_benchmark}
    \end{center}
\end{figure*}  

\subsection{Evaluation on IQA Methods}
\label{sec:iqa:benchmarks}
We select a set of commonly-used IQA methods to build a benchmark.
For the FR-IQA methods, we include: PSNR, NQM \citep{nqm}, UQI \citep{uqi}, SSIM \citep{ssim}, MS-SSIM \citep{ms-ssim}, IFC \citep{ifc}, VIF \citep{vif}, VSNR-FR \citep{vsnr}, RFSIM \citep{rfsim}, GSM \citep{gsm}, SR-SIM \citep{sr-sim}, FSIM and FSIM$_\mathcal{C}$ \citep{fsim}, SFF \citep{sff}, VSI \citep{vsi}, SCQI \citep{scqi}, LPIPS-Alex and -VGG \citep{zhang2018unreasonable}\footnote{In this work, we use the 0.1 version of LPIPS.}, PieAPP \citep{prashnani2018pieapp}, DISTS \citep{dists}, and WaDIQaM \citep{wadiqam}.
We also include some popular NR-IQA methods: NIQE \citep{niqe}, Ma \citep{ma2017learning}, and PI \citep{blau2018perception}.
Among them, PI is derived from a combination of NIQE and Ma.
Note that these NR-IQA methods are designed to measure the intrinsic quality of images, and is different from measuring the perceptual similarity, thus the direct comparison of these methods is unfair.
All these methods are implemented using the released code.

As in many previous works, we evaluate IQA methods mainly using Spearman rank order correlation coefficients (SRCC) \citep{sheikh2006statistical} and Kendall rank order correlation coefficients (KRCC) \citep{kendall1977advanced}.
These two indexes evaluate the monotonicity of methods: whether the scores of high-quality images are higher (or lower) than low-quality images.
We also include the Pearson linear correlation coefficient (PLCC) results, which are often used to evaluate the accuracy of methods.
Before calculating PLCC index, we perform the third-order polynomial nonlinear regression.
The complete numerical results and full discussion will be shown in Appendix \ref{apd:results}.

\begin{figure*}[t]
    \centering
    \setlength{\tabcolsep}{0.0mm}
    \small
    \resizebox{1.0\linewidth}{!}{
    \begin{tabular}{cccccc}
        & PSNR  & SSIM  & FSIM  & LPIPS  & PieAPP \\
        \rotatebox{90}{~~~Traditional SR} &
        \includegraphics[width=0.25\linewidth]{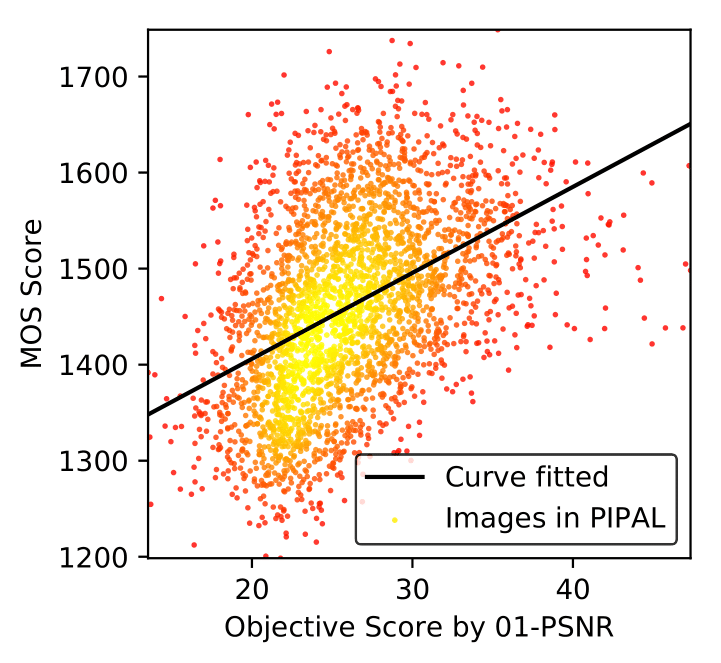} & 
        \includegraphics[width=0.25\linewidth]{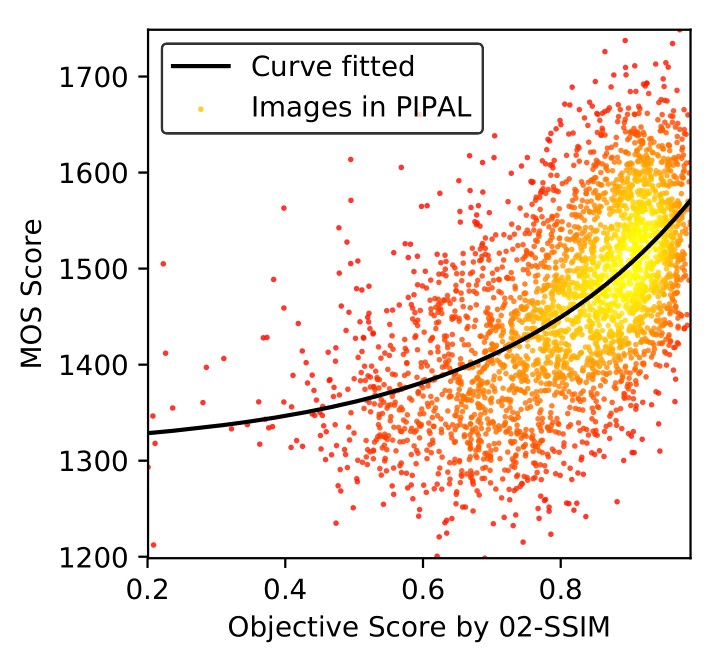} & 
        \includegraphics[width=0.25\linewidth]{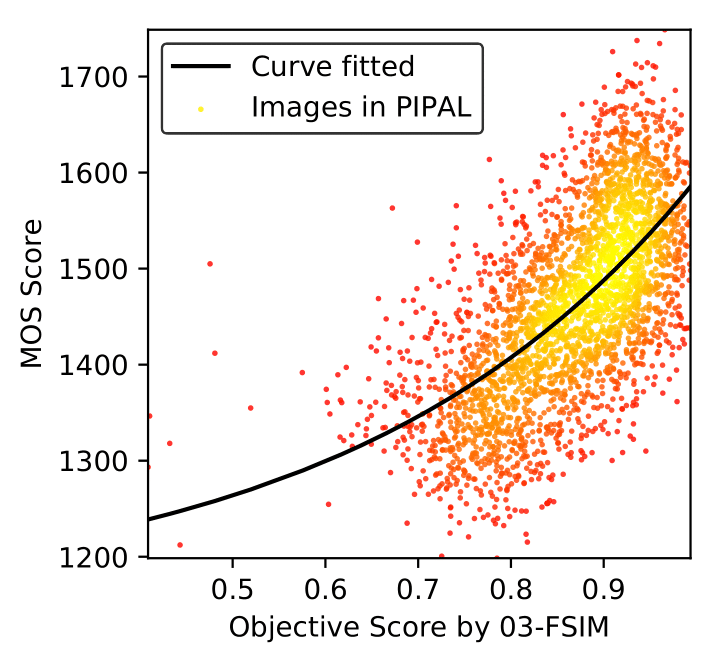} &
        \includegraphics[width=0.25\linewidth]{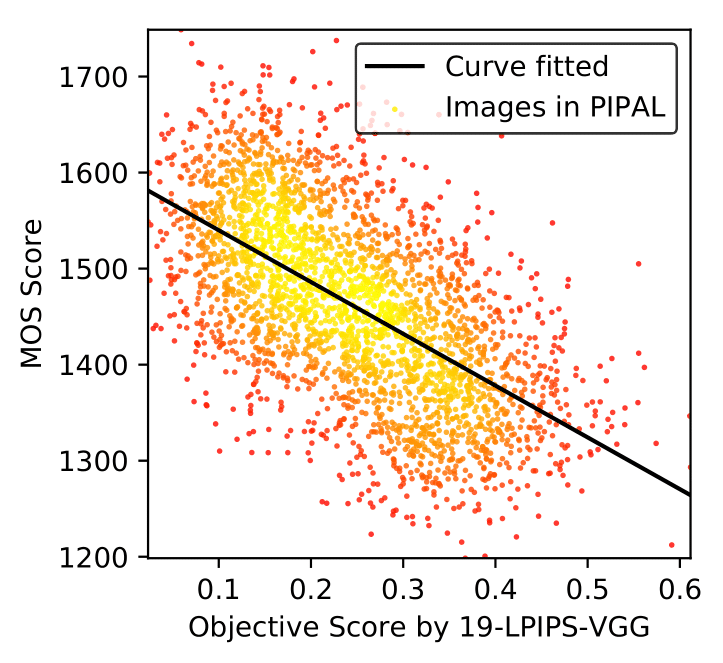} & 
        \includegraphics[width=0.25\linewidth]{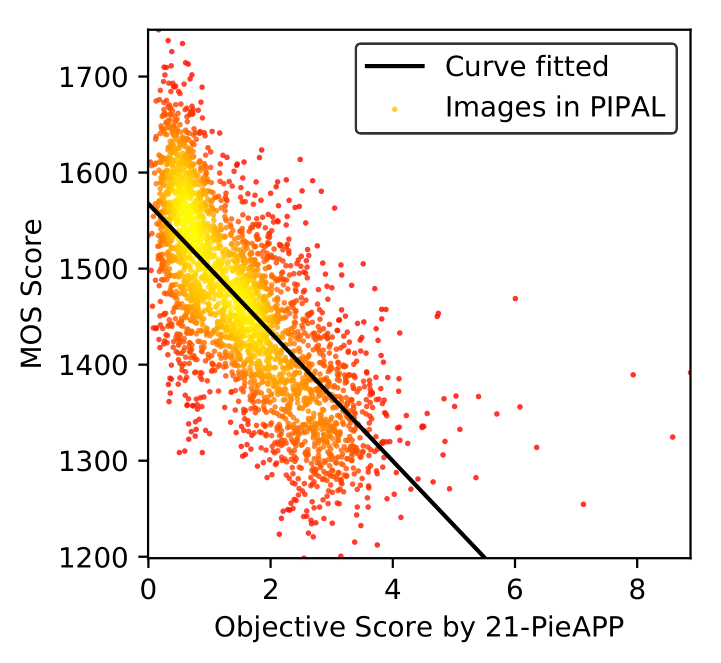}\\
        \rotatebox{90}{PSNR-oriented SR} &
        \includegraphics[width=0.25\linewidth]{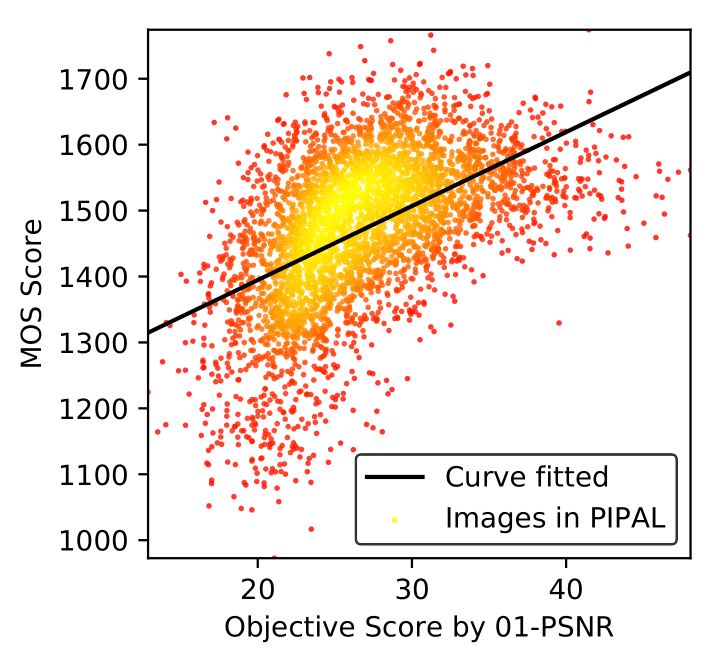} & 
        \includegraphics[width=0.25\linewidth]{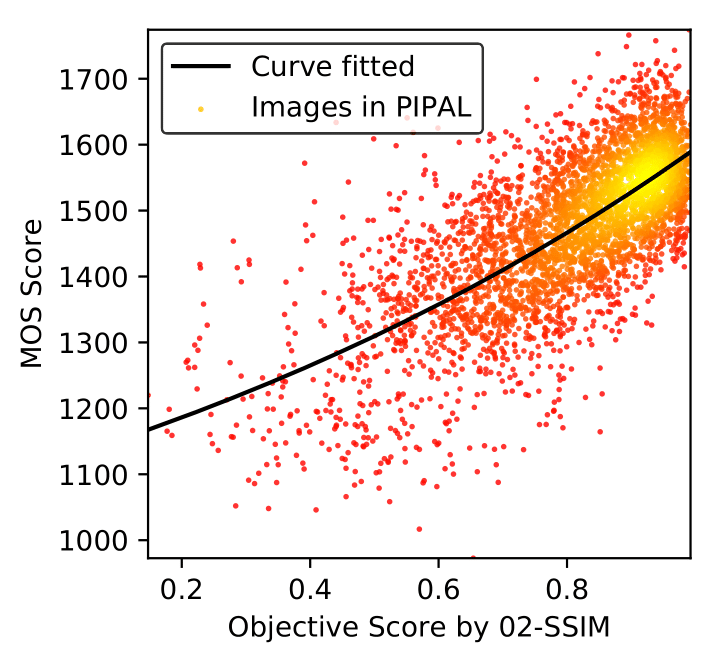} & 
        \includegraphics[width=0.25\linewidth]{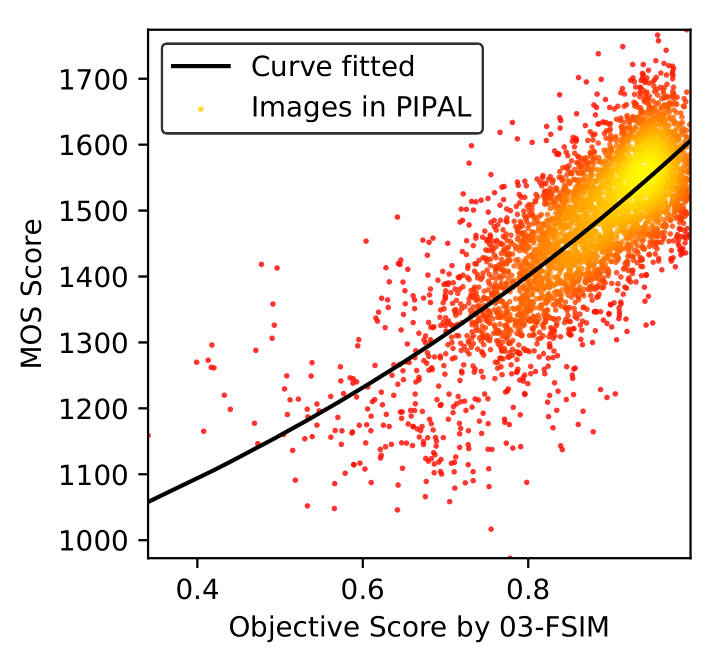} &
        \includegraphics[width=0.25\linewidth]{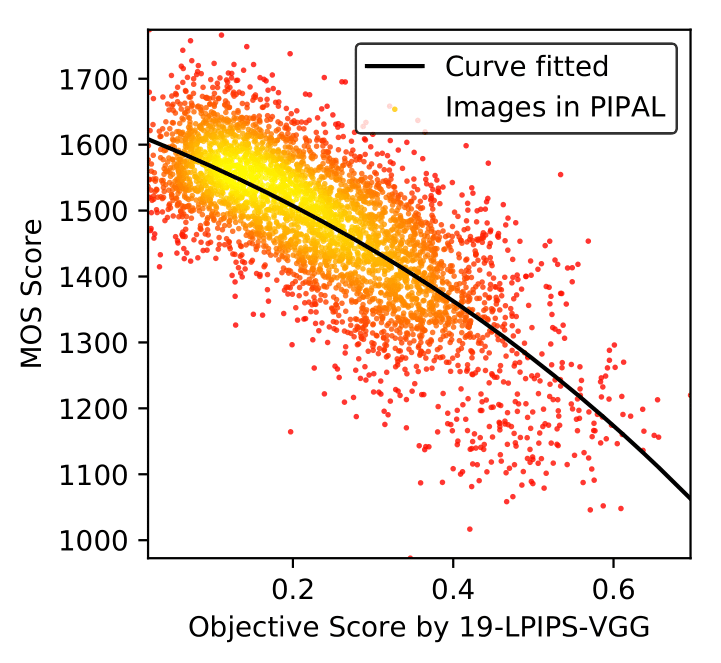} & 
        \includegraphics[width=0.25\linewidth]{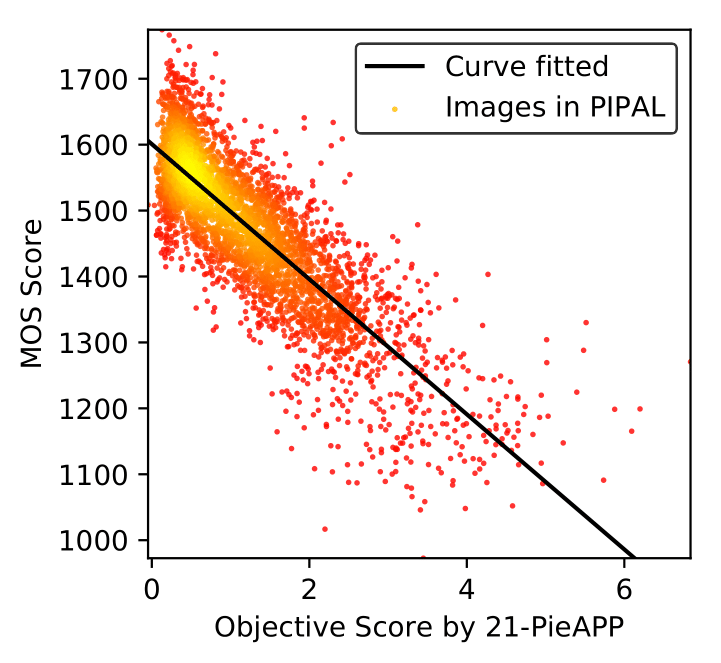}\\
        \rotatebox{90}{~~~GAN-based SR} &
        \includegraphics[width=0.25\linewidth]{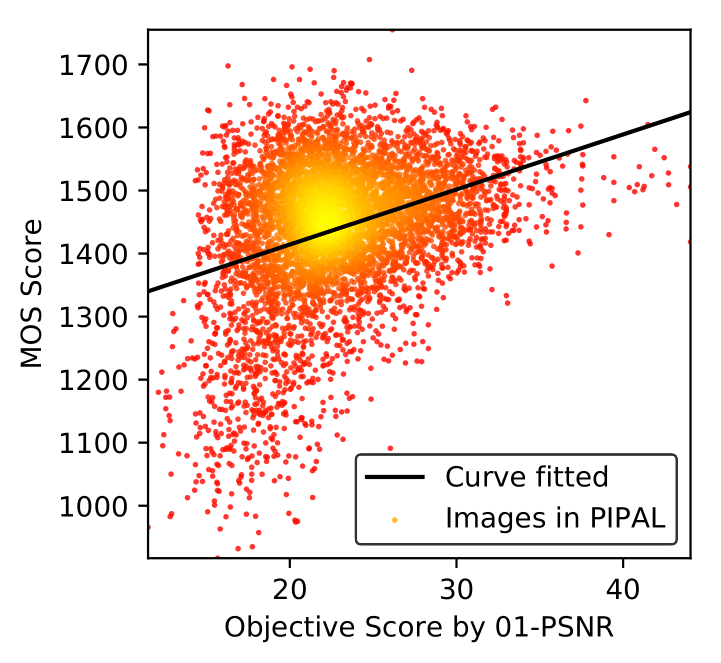} & 
        \includegraphics[width=0.25\linewidth]{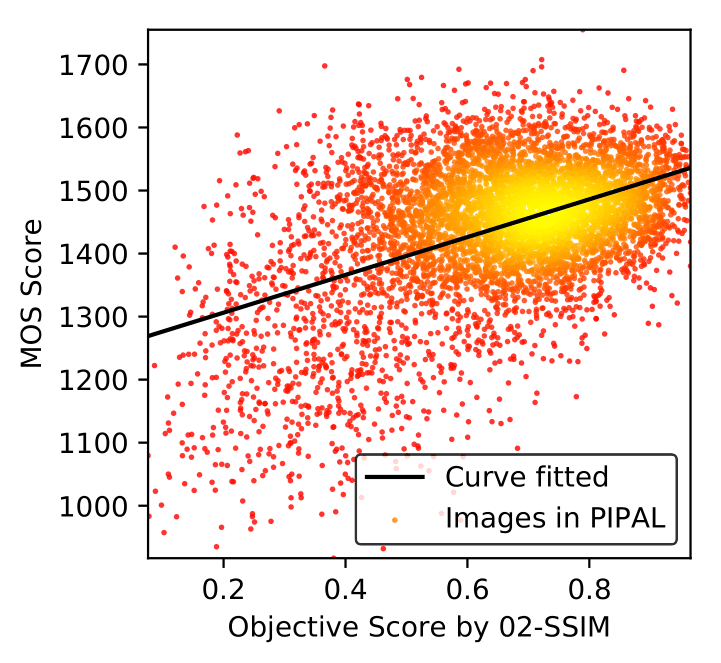} & 
        \includegraphics[width=0.25\linewidth]{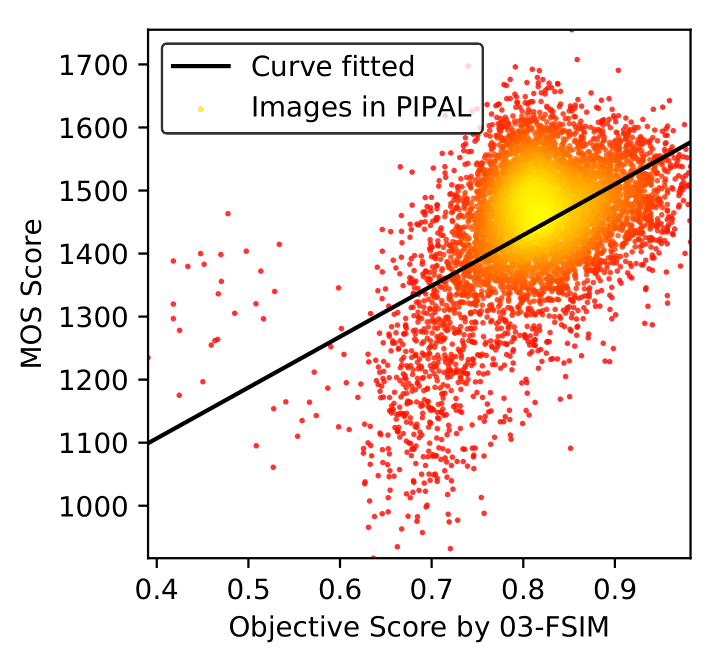} &
        \includegraphics[width=0.25\linewidth]{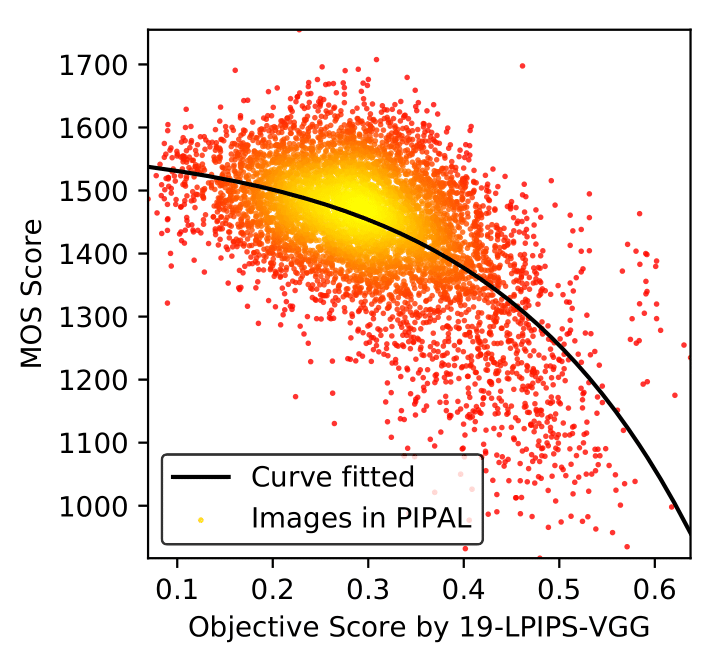} & 
        \includegraphics[width=0.25\linewidth]{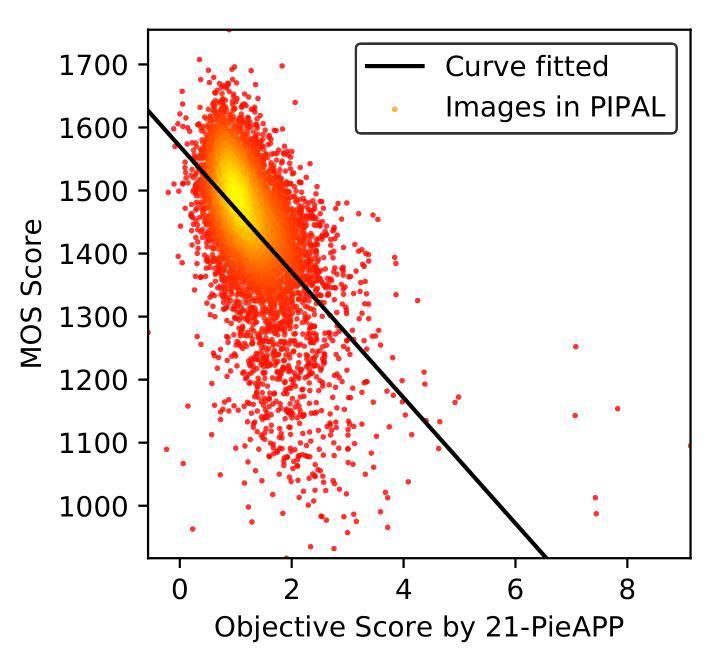}\\
    \end{tabular}
    }
    \caption{Scatter plots of the objective scores vs. the MOS scores for different SR algorithm types. One can see that for Traditional and PSNR-oriented SR, the objective scores predicted by FSIM have a higher correlation with the MOS scores. For GAN-based SR, the performance decline obviously.}
    \label{fig:scatter}
\end{figure*}  

We first evaluate the IQA methods using all types of distortions in PIPAL dataset.
A convenient presentation for both SRCC and KRCC rank coefficients is shown in \figurename~\ref{fig:rcc}.
The first conclusion is that even the best IQA method -- PieAPP provides SRCC with only about 0.71 for all distortion types, which is much lower than its performance on TID2013 dataset (about 0.90).
This indicates that the proposed PIPAL dataset is challenging for existing IQA methods and there is a large room for future improvement.
Moreover, a high overall correlation performance does not necessarily indicate the high performance on each sub-type of distortions.
As the focus of this paper, we want to analyze the performance of IQA using IR results, especially the outputs of GAN-based algorithms.
Specifically, we take SR sub-type as an example and show the performance of IQA methods in evaluating SR algorithm distortions.
In \tablename~\ref{tab:iqa_results}, we show the SRCC results with respect to different distortion sub-types, including traditional distortions, denoising outputs, all SR outputs, traditional SR outputs, PSNR-oriented SR outputs, and GAN-based outputs.
Analysis of \tablename~\ref{tab:iqa_results} leads to the following conclusions.
First, although performing well in evaluating traditional and PSNR-oriented SR algorithms, all of these IQA methods suffer from severe performance drop when evaluating GAN-based distortion.
This confirms the conclusion of \citet{blau2018perception} that higher PSNR may be related to lower perceptual performance for GAN-based algorithms.
Second, despite the severe performance drop, several IQA methods (e.g., LPIPS, PieAPP, and DISTS) still outperform the others.
Coincidentally, these methods are all recent works and based on deep networks.
We also show the scatter distributions of the subjective scores vs. the predicted scores in \figurename~\ref{fig:scatter} on the PIPAL dataset by some representative IQA methods.
The curves shown in \figurename~\ref{fig:scatter} were obtained by the third-order polynomial non-linear fitting.
For the nonlinear regression, we follow the suggestion of Sheikh \textit{et al.} \citep{sheikh2006statistical}.
One can see that for the distortions without GAN effects, the objective scores have a higher correlation with the subjective scores than GAN-based distortion.

We then present the analysis of IQA methods as IR performance measures.
In \figurename~\ref{fig:sr_benchmark}, we show the scatter plots of subjective scores vs. some commonly-used image quality metrics and their correlations for 23 SR algorithms.
Among them, PSNR, SSIM and FSIM are the most common measures, IFC is suggested by \citet{yang2014single} for its good performance on early algorithms, NIQE and PI are suggested in recent works \citep{blau2018perception} for their good performance on GAN-based algorithms.
LPIPS \citep{zhang2018unreasonable} and PieAPP \citep{prashnani2018pieapp} are selected according to our benchmark.
As can be seen that PSNR, SSIM and IFC are \emph{anti-correlated} with the subjective scores, thus are inappropriate for evaluating GAN-based algorithms.
NIQE and PI show moderate performance in evaluating IR algorithms, while LPIPS and PieAPP are the most correlated.
Note that different from the work of \citet{blau2018perception}, where they evaluate the perceptual quality only based on whether the image looks real, we collect subjective scores based on the perceptual similarity with the ground truth.
Therefore, in evaluating the performance of the IR algorithms from the perspective of reconstruction, the suggestions given by our work are more appropriate.
These results have a similar trend as the results presented in \figurename~\ref{fig:scatter}.

\begin{figure*}
    \centering
    \includegraphics[width=1.0\linewidth]{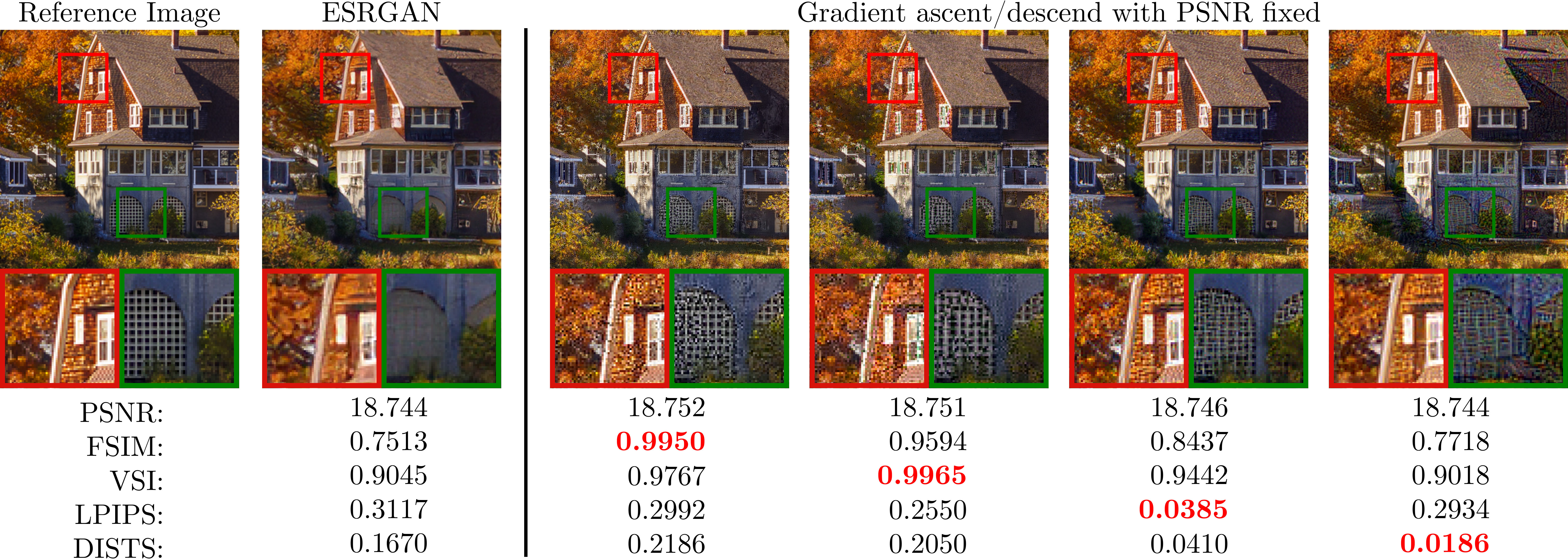}
    \caption{The ``counter-examples`` with respect to different IQA methods with identical PSNR. These are computed by projected gradient ascent/descent optimization on certain IQA methods.}
    \label{fig:optim}
\end{figure*}  

\begin{table*}[t]
\begin{center}
    \footnotesize
    {\resizebox{\linewidth}{!}{
  \begin{tabular}{
  p{2cm}
  p{1.4cm}
  p{1.4cm}
  p{1.4cm}
  p{1.4cm}
  p{1.4cm}
  p{1.4cm}
  p{1.4cm}
  p{1.4cm}
  }
        \toprule
        Method &
        Year &
        PSNR $\uparrow$ &
        ~SSIM $\uparrow$ &
        \underline{Ma} $\uparrow$ &
        ~\underline{NIQE} $\downarrow$ &
        ~~~~\underline{PI} $\downarrow$ &
        LPIPS $\downarrow$ &
        ~~MOS $\uparrow$ \\
        \specialrule{0em}{1pt}{1pt}
        \hline
        \specialrule{0em}{1pt}{1pt}
        YY          & 2013 & 23.35$^8$    & 0.6897$^7$    & 4.5486$^{10}$ & 6.4174$^8$    & 5.9344$^7$    & 0.3574$^{12}$ & 1367.71$^8$ \\
        TSG         & 2013 & 23.55$^7$    & 0.6775$^8$    & 4.1298$^{12}$ & 6.4163$^7$    & 6.1433$^{10}$ & 0.3570$^{11}$ & 1387.24$^7$ \\
        A+          & 2014 & 23.82$^6$    & 0.6919$^6$    & 4.3852$^{11}$ & 6.3645$^5$    & 5.9897$^9$    & 0.3491$^{10}$ & 1354.52$^{12}$ \\
        SRCNN       & 2014 & 23.93$^5$    & 0.6966$^5$    & 4.6094$^9$    & 6.5657$^{10}$ & 5.9781$^8$    & 0.3316$^8$    & 1363.68$^{11}$ \\
        FSRCNN      & 2016 & 24.07$^4$    & 0.7013$^3$    & 4.6686$^8$    & 6.9985$^{11}$ & 6.1649$^{11}$ & 0.3281$^7$    & 1367.49$^9$ \\
        VDSR        & 2016 & 24.13$^3$    & 0.6984$^4$    & 4.7799$^7$    & 7.4436$^{12}$ & 6.3319$^{12}$ & 0.3484$^9$    & 1364.90$^{10}$ \\
        EDSR        & 2017 & \textbf{25.17}$^2$    & \textbf{0.7541}$^2$    & 5.7634$^6$    & 6.4560$^9$    & 5.3463$^6$    & 0.3016$^6$    & 1447.44$^6$ \\
        SRGAN       & 2017 & 22.57$^{10}$ & 0.6494$^{11}$ & 8.4215$^3$    & 3.9527$^3$    & 2.7656$^3$    & \textbf{0.2687}$^2$    & 1494.14$^3$ \\
        RCAN        & 2018 & \textbf{25.21}$^1$    & \textbf{0.7569}$^1$    & 5.9260$^5$    & 6.4121$^6$    & 5.2430$^5$    & 0.2992$^5$    & 1455.31$^5$ \\
        BOE         & 2018 & 22.68$^9$    & 0.6582$^9$    & \textbf{8.5209}$^2$    & \textbf{3.7945}$^1$    & \textbf{2.6368}$^2$    & 0.2933$^4$    & 1481.51$^4$ \\
        ESRGAN      & 2018 & 22.51$^{11}$ & 0.6566$^{10}$ & 8.3424$^4$    & 4.7821$^4$    & 3.2198$^4$    & \textbf{0.2517}$^1$    & \textbf{1534.25}$^1$ \\
        RankSRGAN   & 2019 & 22.11$^{12}$ & 0.6392$^{12}$ & \textbf{8.6882}$^1$    & \textbf{3.8155}$^2$    & \textbf{2.5636}$^1$    & 0.2755$^3$    & \textbf{1518.29}$^2$ \\
        \toprule
    \end{tabular}
    }}
    \caption{The $\times4$ SR results. The years of publication are also provided. $\uparrow$ means the higher the better while  $\downarrow$ means the lower the better. The \textbf{bolded} values are the top 2 values and the superscripts indicate the ranking.
    }
    \label{tab:sr_evaluation}
\end{center}
\end{table*}  

\subsection{Evaluation on IR Methods}
\label{sec:ir:benchmarks}
One of the most important applications of IQA technology is to evaluate IR algorithms.
IQA methods have been the chief reason for the progress in the IR field as a means of comparing the performance.
However, evaluating IR methods only with specific IQA methods also narrows the focus of IR research and converts it to competitions only on the quantitative numbers (e.g., PSNR competitions \citep{timofte2018ntire,cai2019ntire} and PI competition \citep{blau20182018}).
As stated above, existing IQA methods may be inadequate in evaluating IR algorithms.
We wonder that with the focus on beating benchmarks on the flawed IQA methods, are we getting better IR algorithms?
To answer this question, we take the SR task as a representative and select 12 SR algorithms to build a benchmark.
These algorithms are all representative algorithms and selected from 2013 to the present.
We evaluate: YY \citep{yy2013}, TSG \citep{tsg2013}, A+ \citep{a+2014}, SRCNN \citep{srcnn2014}, FSRCNN \citep{fsrcnn2016}, VDSR \citep{vdsr2016}, EDSR \citep{edsr2017}, SRGAN \citep{srgan2017}, RCAN \citep{rcan2018}, BOE \citep{boe2018}, ESRGAN \citep{wang2018esrgan}, and RankSRGAN \citep{zhang2019ranksrgan}.
The results are shown in \tablename~\ref{tab:sr_evaluation}.
We present more algorithms in Appendix \ref{apd:results}.
One can observe that before 2017 (when GAN was applied to SR) the PSNR performance improves continuously.
Especially, the deep-learning-based algorithms improve PSNR by about 1.4dB.
These efforts do improve the subjective performance -- the average MOS value increases by about 90 in 4 years.
After SRGAN was proposed, the PSNR decreased by about 2.6dB compared to the state-of-the-art PSNR performance at that time (EDSR), but the MOS value increased by about 50 suddenly.
In contrast, RCAN was proposed to defeat EDSR in terms of PSNR.
Its PSNR performance is only a little higher than EDSR but its MOS score is even lower than EDSR.
When noticing that the mainstream metrics (PSNR and SSIM) had conflicted with the subjective performance, PI was proposed to evaluate perceptual SR algorithms \citep{blau2018perception}.
After that, ESRGAN and RankSRGAN have been continuously improving PI performance.
Among them, the latest RankSRGAN has achieved the current state-of-the-art in terms of PI performance.
However, it is worth noting that, ESRGAN has the highest subjective score, but it has no advantage in the performance of PI and NIQE comparing with RankSRGAN.
Efforts on improving the PI value show limited effects and have failed to continuously improve MOS performance after ESRGAN.
These observations inspire us in two aspects:
(1) None of the existing IQA methods is always effective in evaluation. With the development of IR technology, new IQA methods need to be proposed accordingly.
(2) Excessively optimizing performance on a specific IQA metric may cause a decrease in perceptual quality.

We conduct experiments to explore the outcomes of excessively optimizing IR algorithm performance on an IQA metric by generating ``counter-examples''.
Conceptually speaking, even one counter-example is sufficient to disprove an IQA method as an IR metric, because algorithms may take advantage of this vulnerability to achieve numerical superiority.
We obtain these examples by gradient-based optimization to maximize the values of certain IQA methods.
According to \citet{blau2018perception}, distortion and perceptual quality are at odds with each other.
In order to simulate the situation where there is a perception-distortion trade-off, we constrain the PSNR values to be 
less than or equal to that of the initial distorted image during optimization.
Given a reference image $I_R$ and an initial image $I_i$, we solve the following objective function to obtain the possible ``counter-example'' $x$ for an differentiable IQA methods $\mathcal{H}$:
\begin{align}
    \max_{x} \quad & \mathcal{H}(x,I_R),\\
    \mbox{s.t.}\quad & \|x-I_R\|_2^2\leq a,
    \label{eq:cons}
\end{align}
where $a=\|I_i-I_R\|_2^2$ and this constraint ensures the PSNR value of the result will not increase.
We employ the projected gradient method \citep{boyd2004convex} to solve this optimization problem.
Its basic idea is to take a step in the descending direction, and then project it to the feasible region.
Because the constraint is convex, the projection to the convex set is unique.
The specific iteration formula is as follows:
\begin{align}
    \mathrm{Gradient\ descent:}\quad& y^{k+1}=x^k-\alpha\nabla_x \mathcal{H}(x^k,I_R),\\
    \mathrm{Projection:}\quad&x^{k+1}=\mathop{\arg\min}_{x\in C} \|x-y^{k+1}\|_2^2,
\end{align}
where $C$ is the feasible region that satisfy Eq. \eqref{eq:cons} and $\alpha$ is the learning rate.
In our experiment, We use the output of ESRGAN as the initial image.
Note that for some IQA methods, the higher value indicates the better perceptual quality, thus we perform gradient ascent instead of gradient descent to achieve better quality.

The results are shown in \figurename~\ref{fig:optim}.
We can see that some images show superior numerical performance when evaluated using certain IQA methods, but may not be dominant in other metrics.
Their best-cases also show different visual effects.
Even for some IQA methods (LPIPS and DISTS) with good performance on GAN-based distortion, their best-cases still contain serious artifacts.
This shows the risk of evaluating IR algorithms using these IQA methods.
This also indicates that evaluating and developing new IQA methods plays an important role in future research.

\begin{figure}[!t]
	\begin{center}
		\subfigure[]{\includegraphics[width=0.48\linewidth]{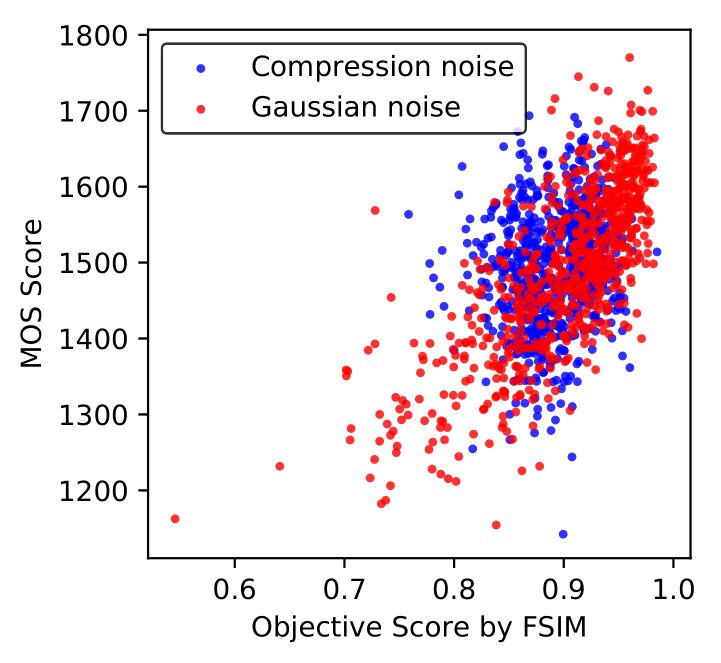}}
		\hspace{0.00cm}
		\subfigure[]{\includegraphics[width=0.48\linewidth]{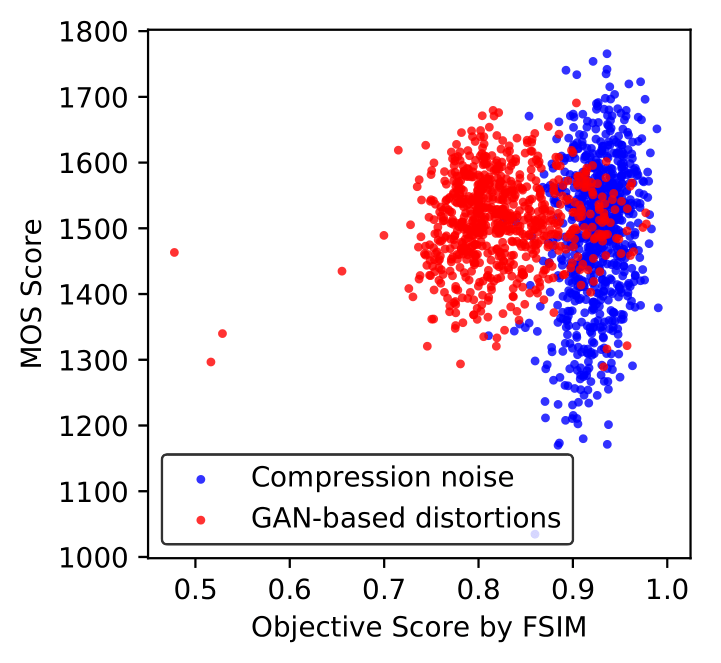}}

		\subfigure[]{\includegraphics[width=0.48\linewidth]{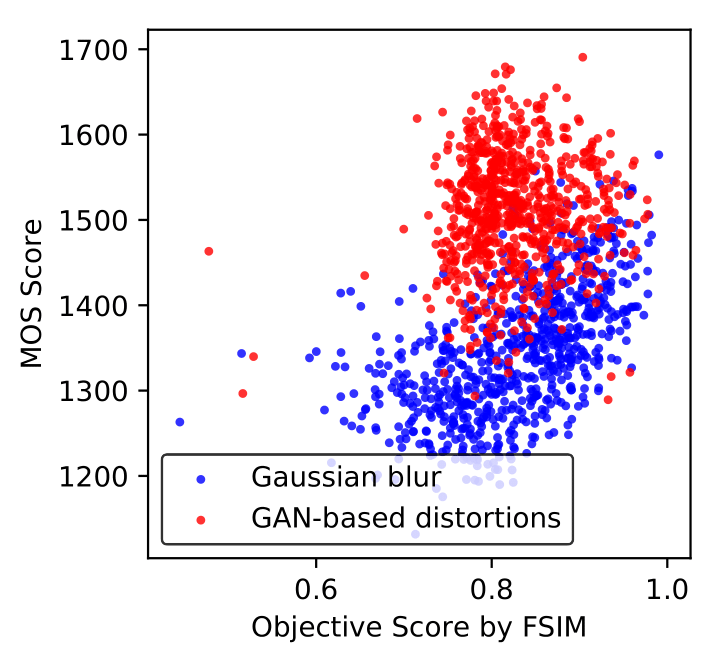}}
		\hspace{0.00cm}
		\subfigure[]{\includegraphics[width=0.48\linewidth]{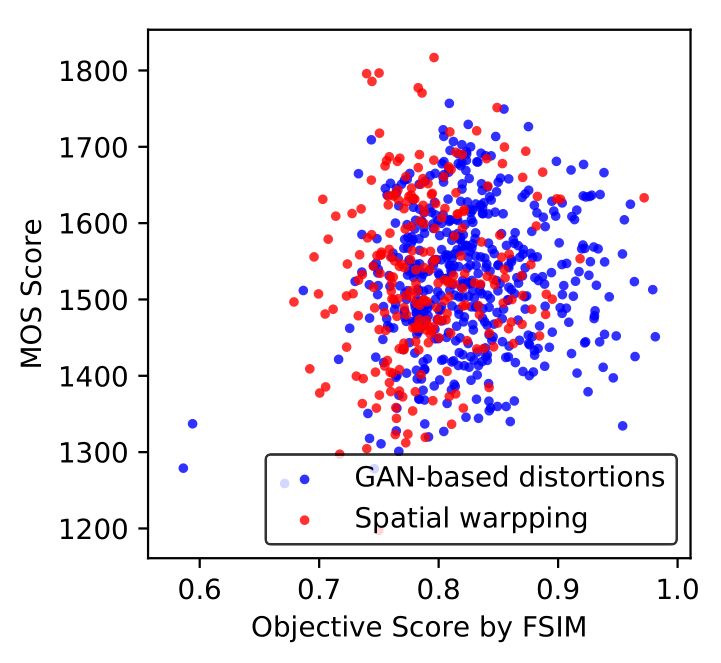}}
		\caption{Examples of scatter plots for pairs of distortion types. For distortion types that are easy to measure, samples are well clustered along the fitted curve. For others that are difficult for the IQA method, the samples will not be well clustered. The samples of distortion types that have similar behavior will overlap with each other.}
		\label{fig:gan_warp}
	\end{center}
\end{figure} 

\subsection{Discussion of GAN-based distortion}
\label{sec:GANDistort}
Recall that LPIPS, DISTS and PieAPP perform relatively better in evaluating GAN-based distortion. 
The effectiveness of these methods may be attributed to the following reasons.
Compared with other IQA methods, deep-learning-based IQA methods can extract image features more effectively.
For traditional distortion types, such as blur, compression and noise, the distorted images usually disobey the prior distribution of natural images.
Early methods can assess these images by measuring low-level statistic features such as image contrast, gradient, and structural information.
These strategies are also effective for the outputs of traditional and PSNR-oriented algorithms.
However, most of these strategies fail in the case of GAN-based distortion, as such distortions may have similar image statistic features with the reference image -- the way that GAN-based distortion differs from the reference image is less apparent.
In this case, deep networks can capture these unapparent features and distinguish such distortions to some extent.

In order to explore the characteristics and difficulties of GAN-based distortion, we compare them with some well-studied distortions.
Note that, for a good IQA method, the subjective scores in the scatter plot should increase coincide with objective values, and the samples are well clustered along the fitted curve.
In the case of two distortion types, if the IQA method behaves similarly for both of them, their samples on the scatter plot will also be well clustered and overlapped.
For example, the additive Gaussian noise and image lossy compression are well studied for most IQA methods.
When calculating the objective values using FSIM$_\mathcal{C}$, samples of both distortions are clustered, as shown in \figurename~\ref{fig:gan_warp} (a).
This indicates that FSIM$_\mathcal{C}$ can adequately characterize the visual quality of an image damaged due to these two types of distortion.
Then we study GAN-based distortion by comparing it with some existing distortion types using FSIM$_\mathcal{C}$.
\figurename~\ref{fig:gan_warp} (b) shows the results of GAN-based distortion and additive Gaussian noise, and \figurename~\ref{fig:gan_warp} (c) shows the results of GAN-based distortion and Gaussian blur.
It can be seen that the samples of Gaussian noise and Gaussian blur barely intersect with GAN-based samples.
FSIM$_\mathcal{C}$ largely underestimates the visual quality of GAN-based distortion.
In \figurename~\ref{fig:gan_warp} (d), we show the result of GAN-based distortion and spatial warping distortion.
As can be seen, these two distortion types behave unexpectedly similar.
FSIM$_\mathcal{C}$ cannot handle them and shows the same random and diffused state.
The quantitative results also verify this phenomenon.
For spatial warping distortion type, the SRCC of FSIM$_\mathcal{C}$ is 0.3070, and it is close to the performance of GAN-based distortion, which is 0.4047.
Thus we argue that the spatial warping distortion and GAN-based distortion pose similar challenges to FSIM$_\mathcal{C}$.

As revealed in experimental psychology \citep{kolers1962intensity,kahneman1968method}, the interaction or mutual interference between visual information may cause the Visual Masking effects.
According to this theory, some key reasons that IQA methods tend to underestimate both GAN-based distortion and spatial warping distortion are as follows.
Firstly, for the edges with strong intensity change, the human visual system (HVS) is sensitive to the contour and shape, but not sensitive to the error and misalignment of the edges.
Secondly, the ability of HVS to distinguish texture decreases in the region with dense textures.
When the extracted features of the textures are similar, the HVS will ignore part of the subtle differences and misalignment of textures.
However, both the traditional and deep-learning-based IQA methods require a good alignment for the inputs.
This partially causes the drop of performance of these IQA methods on GAN-based distortion.
This finding provides an insight that if we explicitly consider the spatial misalignment of the inputs, we may improve the performance on GAN-based distortion.
We will discuss this possibility in Sec~\ref{sec:new_iqa}.

\begin{figure}[t]
    \centering
    \includegraphics[width=1.0\linewidth]{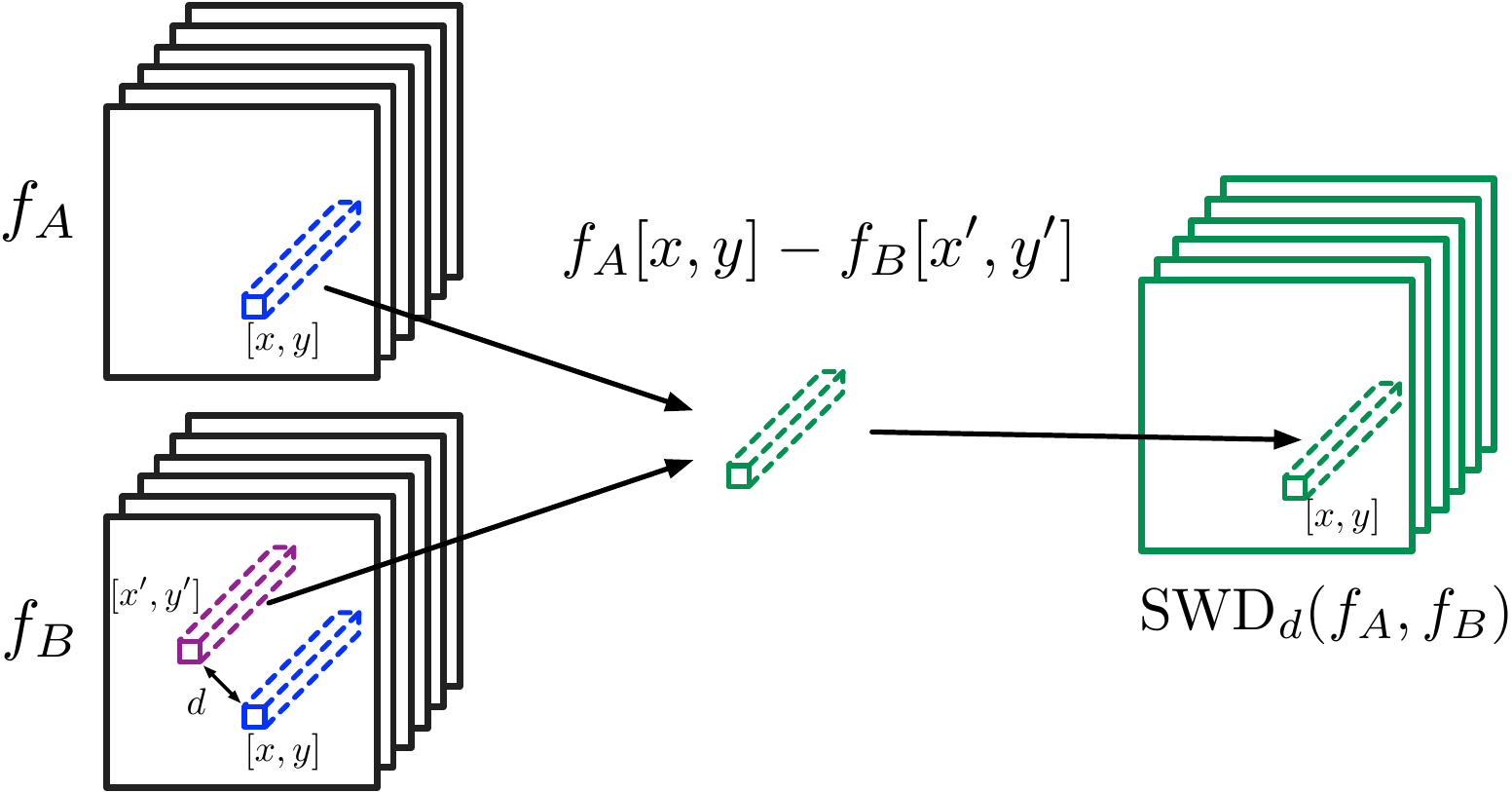}
    \caption{
    The proposed Space Warping Difference (SWD) layer.
    We compare the features that not only on the corresponding position but also on a small range around the corresponding position to explicitly consider the robustness of spatial misalignment.}
    \label{fig:SWD}
\end{figure}

\begin{figure*}[t]
    \centering
    \includegraphics[width=1.0\linewidth]{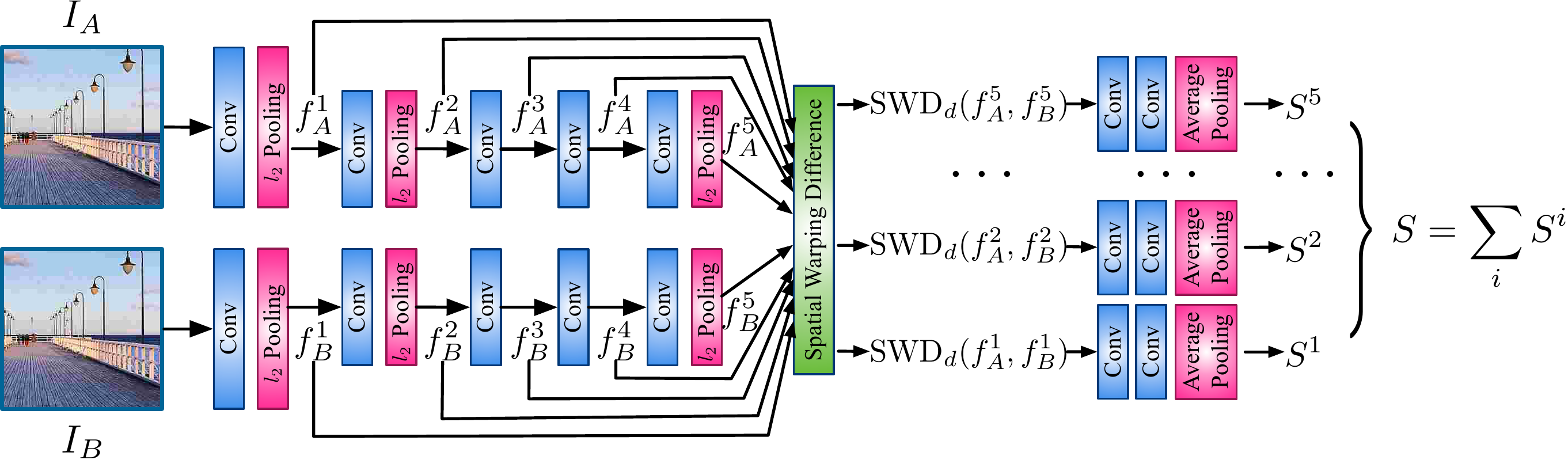}
    \caption{The network architecture of the proposed Space Warping Difference IQA Network (SWDN).}
    \label{fig:SWPS}
\end{figure*}  

\section{Improving IQA Networks}
\label{sec:new_iqa}
As stated in Sec~\ref{sec:GANDistort}, we argue that deep-learning-based methods outperform others due to their excellent feature extraction capacity.
However, when it comes to the distortions with spatial misalignment such as GAN-based distortion, these methods obtain unsatisfactory performance, partially because of their low tolerance to misalignment.
This finding provides us an insight that if we explicitly consider the spatial misalignment, we may improve the performance of IQA methods on GAN-based distortion.
In this section, we explore this possibility by introducing anti-aliasing pooling layer and a spatially robust comparison operation into the IQA network.

\subsection{Anti-aliasing Pooling and Space Warping Difference}
\label{sec:swdn}
FR-IQA networks can be roughly divided into three sub-models: feature extraction, feature comparison and subjective score regression.
Among them, the feature extraction sub-model extracts image features by cascaded convolution operations.
Feature comparison sub-model compares the features of two images.
The most direct and commonly used method is to calculate the differences of image feature vectors \citep{zhang2018unreasonable}.
Finally, the subjective score regression sub-model calculates the final similarity scores based on the feature differences.
If we want the IQA networks to be robust to small misalignment, the feature extraction and feature comparison sub-models should at least be invariant to this misalignment/shift.
However, these two parts are usually not shift-invariant in existing deep IQA networks.
We then discuss the feature extraction and feature comparison sub-models, respectively.

For the feature extraction sub-model, the standard convolution operations should have been shift-invariant.
However, some commonly used downsampling layers, such as max/average pooling and strided-convolution, ignore the sampling theorem \citep{azulay2019deep} and are not shift-invariant anymore.
These downsampling layers are widely used in deep networks such as VGG \citep{simonyan2014very} and AlexNet \citep{krizhevsky2012imagenet}, which are popular backbone architectures for feature extraction (e.g., in LPIPS and DISTS).
As suggested by \citet{zhang2019making}, one can fix this by introducing anti-aliasing pooling.
In our work, following \citet{henaff2016geodesics}, to avoid aliasing when downsampling by a factor of two, the Nyquist theorem requires blurring with a filter whose cutoff frequency is below $\frac{\pi}{2}$.
Then, we introduce the $l_2$ pooling:
\begin{equation}
    P(x)=\sqrt{g\otimes (x \odot x)},
\end{equation}
where $\odot$ represents the Hadamard product, $\otimes$ represent convolution operation and $g$ is the blurring kernel and is implemented by a Hanning window that approximately enforces the Nyquist criterion.
In this work, we replace all the max pooling layers with $l_2$ pooling layers in our IQA network to avoid possible aliasing in feature extraction.

We then discuss the feature comparison sub-model.
In most existing IQA networks, the comparison is conducted with element-wise subtraction or a Euclidean distance between two extracted features.
These operations require good alignment and are all sensitive to spatial shift.
Here, we explicitly consider the robustness of spatial misalignment and propose a Space Warping Difference (SWD) layer to compare the features that not only on the corresponding position but also on a small range around the corresponding position.
The SWD layer is illustrated in \figurename~\ref{fig:SWD}.
For two images $I_A$ and $I_B$ and their extracted features $f_A$ and $f_B$, the SWD layer is formulated as follow:
\begin{equation}
    \mathrm{SWD}_d(f_A,f_B)[x,y]=f_A[x,y]-f_B[x',y'],
\end{equation}
where $d$ indicates the searching range, $f[x,y]$ indicates the feature vector at location $(x,y)$, and $(x',y')$ indicates the location that the $l_2$ distance of two feature vectors achieves the minimum value:
\begin{align}
    &x',y'=\mathop{\arg\min}_{x',y'} \|f_A[x,y]-f_B[x',y']\|_2^2,\\
    &s.t. \quad |x'-x|\leq d, |y'-y|\leq d.
\end{align}
Because a big $d$ will make the computational complexity rise sharply, in our work, we choose $d=3$ for a better performance speed trade-off.

\begin{figure}[t]
    \centering
    \includegraphics[width=1.0\linewidth]{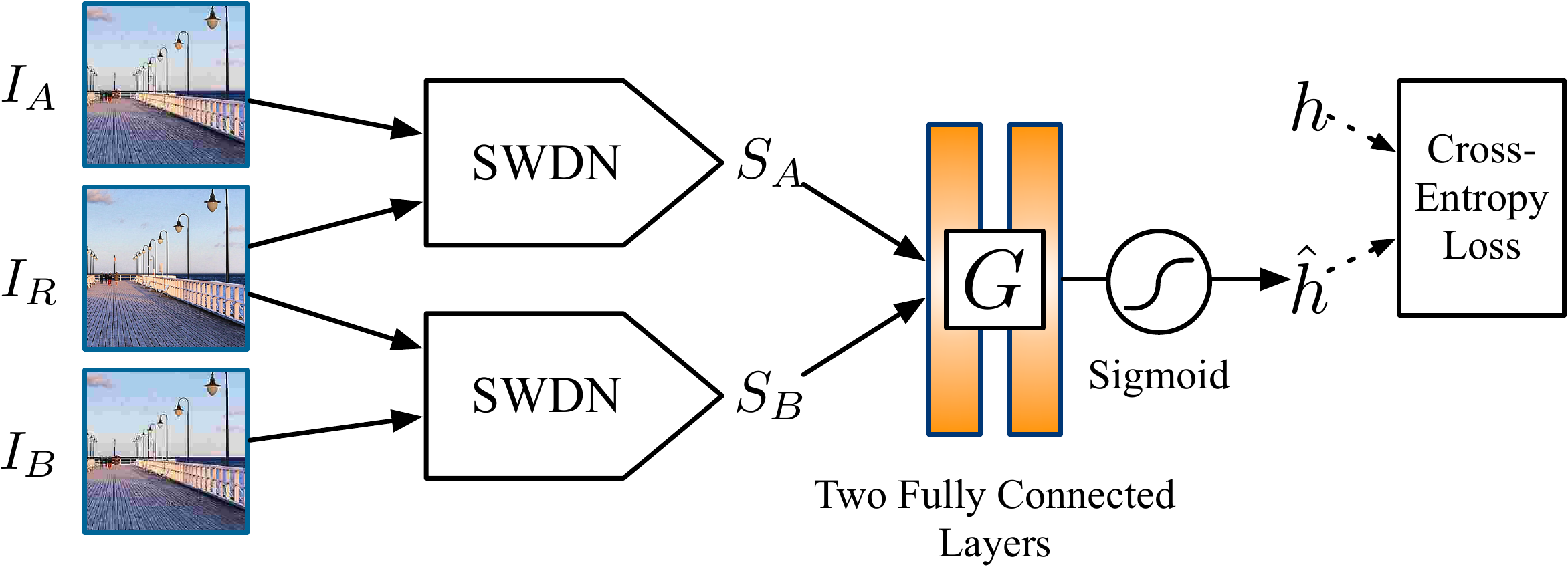}
    \caption{
    The training framework of the proposed SWDN.
    We first calculate the similarity scores between the reference image $I_R$ and $I_A$ and $I_B$.
    And then a small network $G$ is trained to predict preference probability  $\hat{h}$ from score pair $(S_A,S_B)$.
    Finally, we compute the loss using cross entropy loss function.
    }
    \label{fig:training}
\end{figure}  

\begin{table*}[t]
    \centering
    \footnotesize
    \resizebox{1.0\linewidth}{!}{
    \begin{tabular}{l
    p{1.2cm}<{\centering}p{1.2cm}<{\centering}
    p{1.2cm}<{\centering}p{1.2cm}<{\centering}
    p{1.2cm}<{\centering}p{1.2cm}<{\centering}
    p{1.2cm}<{\centering}p{1.2cm}<{\centering}
    p{1.2cm}<{\centering}p{1.2cm}<{\centering}}
    \multirow{2}{*}{Method} &
    \multicolumn{2}{c}{PIPAL (full set)} & \multicolumn{2}{c}{TID2013} & \multicolumn{2}{c}{Trad. \& PSNR. SR} & \multicolumn{2}{c}{GAN distort.} & \multicolumn{2}{c}{PIPAL (test set)} \\
    & SRCC & KRCC & SRCC & KRCC & SRCC & KRCC & SRCC & KRCC & SRCC & KRCC \\
    
    \hline
    PSNR &
    0.4073 & 0.2771 & 0.6870 & 0.4960 & 0.5227 & 0.3577 & 0.2839 & 0.1939  & 0.4262 & 0.2931 \\
    SSIM &
    0.5285 & 0.3689 & 0.7417 & 0.5588 & 0.6510 & 0.4588 & 0.3388 & 0.2314  & 0.5277 & 0.3709\\
    FSIM &
    0.5961 & 0.4214 & 0.8015 & 0.6289 & \textbf{0.7083} & \textbf{0.5080} & 0.4090 & 0.2806  & 0.6046 & 0.4284 \\
    \underline{NIQE} &
    0.0131 & 0.0086 & 0.2770 & 0.1844 & 0.1084 & 0.0721 & 0.0155 & 0.0102  & 0.0861 & 0.0555\\
    \underline{Ma \textit{et al.}} &
    0.4216 & 0.2951 & 0.3496 & 0.2417 & 0.6774 & 0.4863 & 0.0545 & 0.0363  & 0.2843 & 0.2040\\
    PieAPP &
    \textbf{0.7113} & \textbf{0.5212} & \textbf{0.9450} & \textbf{0.8040} & \textbf{0.7828} & \textbf{0.5813} & \textbf{0.5530} & \textbf{0.3865}  & \textbf{0.6589} & \textbf{0.4797}\\
    DISTS &
    \textbf{0.6639} & \textbf{0.4774} & \textbf{0.8300} & \textbf{0.6390} & \textbf{0.7348} & \textbf{0.5321} & \textbf{0.5527} & \textbf{0.3839}  & 0.6324 & 0.4447\\
    LPIPS-VGG  &
    0.5772 & 0.4075 & 0.6695 & 0.4969 & 0.6839 & 0.4865 & 0.4816 & 0.3329  & 0.6006 & 0.4265\\
    LPIPS-Alex &
    0.5873 & 0.4116 & 0.7444 & 0.5477 & 0.6293 & 0.4397 & 0.4882 & 0.3322  & 0.5332 & 0.3645  \\
    SWDN (ours) &
    \textbf{0.6530} & \textbf{0.4682} & 0.7895 & 0.6065 & 0.6631 & 0.4706 & \textbf{0.5512} & \textbf{0.3859}  & 0.6319 & 0.4502\\
    \hdashline
    LPIPS-Alex* &
    \textcolor[rgb]{0.4,0.4,0.4}{0.7006} & 
    \textcolor[rgb]{0.4,0.4,0.4}{0.5113} & 
    0.7604 & 0.5800 & 
    \textcolor[rgb]{0.4,0.4,0.4}{0.7316} & 
    \textcolor[rgb]{0.4,0.4,0.4}{0.5298} & 
    \textcolor[rgb]{0.4,0.4,0.4}{0.5947} & 
    \textcolor[rgb]{0.4,0.4,0.4}{0.4203} & 
    \textbf{0.6841} & \textbf{0.4947}  \\
    SWDN* (ours)&
    \textcolor[rgb]{0.4,0.4,0.4}{\textbf{0.7301}} & \textcolor[rgb]{0.4,0.4,0.4}{\textbf{0.5383}} & 
    \textbf{0.8187} & \textbf{0.6335} & 
    \textcolor[rgb]{0.4,0.4,0.4}{\textbf{0.7631}} & 
    \textcolor[rgb]{0.4,0.4,0.4}{\textbf{0.5615}} & 
    \textcolor[rgb]{0.4,0.4,0.4}{\textbf{0.6024}} & 
    \textcolor[rgb]{0.4,0.4,0.4}{\textbf{0.4265}} & 
    \textbf{0.7029} & \textbf{0.5114}\\
    \toprule
    \end{tabular}
    }
    \caption{Quantitative Comparison.
    We compare the proposed SWDN methods with the state-of-the-art IQA methods on both TID2013 dataset and the proposed PIPAL dataset.
    The methods marked with ``*'' are trained using the proposed PIPAL training set.
    Training with PIPAL training set have an advantage when testing on itself. These values are indicated by \textcolor[rgb]{0.4,0.4,0.4}{gray} values.
    The \textbf{bloded} values are the highest three numbers. The best gray value per column is also \textcolor[rgb]{0.4,0.4,0.4}{\textbf{bolded}}.
    }
    \label{tab:iqa_net}
\end{table*}  

The proposed network architecture is illustrated in \figurename~\ref{fig:SWPS} and is termed as Space Warping Difference IQA Network (SWDN).
As stated by \citet{zhang2018unreasonable}, features of the classification networks are useful for building perceptual metrics.
We employ an AlexNet pre-trained on ImageNet \citep{russakovsky2015imagenet} as the feature extraction backbone and replace all the max pooling layers with the proposed $l_2$ pooling layers.
We obtain features $f^i$, $i\in\{1,2,\dots,5\}$ after each convolution layer for both $I_A$ and $I_B$ using the feature extraction sub-model.
Note that the feature extraction backbone architecture is replaceable.
Alternative choices include VGG, SqueezeNet \citep{simonyan2014very}, and so on. %
For each pair of extracted features, we use the SWD layer to calculate the difference $\mathrm{SWD}_d(f_A^i,f_B^i)$.
We then perform perceptual score regression by a small sub-net-work with two convolution layers.
This sub-network takes the feature difference as input and gives predicted similarity score $S^i$ for each feature pair.
Finally, we sum up all the $S^i$s to obtain the final perceptual similarity $S$.

We train our network using BAPPS \citep{zhang2018unreasonable} dataset.
We employ ranking cross entropy loss to train the SWDN network.
The training procedure is illustrated in \figurename~\ref{fig:training}.
For each step, we randomly choose one reference image $I_R$ and its two distorted images $I_A$ and $I_B$.
We calculate the distances between the reference image and the distorted images $(S_A,S_B)$.
Given these two distances, we train a small network $G$ on top to map to a probability $\hat{h}\in(0,1)$, where $\hat{h}$ represents the predicted preference between $I_A$ and $I_B$ and $h$ is the ground truth preference probability provided in BAPPS dataset.
The architecture of $G$ consist of two 32-unit fully connected layers with ReLU activations, followed by a 1-unit fully connected layer and a sigmoid activation.
The final loss function is shown as follow:
\begin{align}
    \mathcal{L}(S_A,S_B,h)=&-h\log G(S_A, S_B)\\&-(1-h)\log(1-G(S_A,S_B)).
\end{align}
We note that the proposed PIPAL dataset can also be used to train our network.
According to the Elo scores of $I_A$ and $I_B$, the ground truth preference probability $h$ is obtained using Eq. \eqref{eq:elo}.
For optimization, we use Adam \citep{kingma2014adam} with $\beta_1=0.9$, $\beta_2=0.999$ and learning rate $1\times10^{-4}$.
We implement our models with the Pytorch framework \citep{pytorch} and the whole training process takes about 12 hours.

\begin{table}[t]
    \centering
    \resizebox{1.0\linewidth}{!}{
    \begin{tabular}{
    p{.4cm}<{\centering}
    p{1.2cm}<{\centering}
    p{1.4cm}<{\centering}
    p{1.cm}<{\centering}p{1.cm}<{\centering}
    p{1.cm}<{\centering}p{1.cm}<{\centering}}
        \toprule
        \multirow{2}{*}{No.} & \multirow{2}{*}{$l_2$ Pooling} & \multirow{2}{*}{SWD Layer} &
        \multicolumn{2}{c}{PIPAL (full set)} & \multicolumn{2}{c}{PIPAL (GAN)} \\
        
         & & & SRCC & KRCC & SRCC & KRCC \\
         \hline
        1 & &                               & 0.5604 & 0.3910 & 0.4862 & 0.3339  \\ 
        2 & $\checkmark$ &                  & 0.5816 & 0.4080 & 0.4942 & 0.3394 \\ 
        3 & & $\checkmark,d=1$              & 0.5918 & 0.4160 & 0.5135 & 0.3549 \\ 
        4 & & $\checkmark,d=2$              & 0.6469 & 0.4638 & 0.5478 & 0.3839 \\ 
        5 & & $\checkmark,d=3$              & \textbf{0.6535} & \textbf{0.4691} & \textit{0.5504} & \textbf{0.3859} \\ 
        6 & & $\checkmark,d=4$              & 0.6516 & 0.4674 & 0.5481 & \textit{0.3840} \\ 
        7 & & $\checkmark,d=5$              & 0.6466 & 0.4634 & 0.5438 & 0.3806 \\ 
        8 & $\checkmark$ & $\checkmark,d=3$ & \textit{0.6530} & \textit{0.4682} & \textbf{0.5512} & \textbf{0.3859} \\ 
        \toprule
    \end{tabular}}
    \caption{Ablation study of the proposed components.
    For the baseline architecture, we replace the $l_2$ pooling layers with the original max pooling layers and remove SWD layers.
    We indicate the best performance with \textbf{blod} values and the second best with \textit{italic} values.}
    \label{tab:SWD_ablation}
\end{table}  

\subsection{Results}
\label{sec:iqanet:exp}
In this section, we experimentally verify the effectiveness of the proposed SWDN network.
We provide a comparison with some commonly-used IQA methods, including PSNR, SSIM, FSIM, NIQE, Ma, PieAPP, DISTS, LPIPS-VGG, and LPIPS-Alex.
We test these IQA methods using the proposed PIPAL dataset.
We also provide the results tested on the TID2013 dataset to show their performance on the traditional distortion types.
The results are shown in \tablename~\ref{tab:iqa_net}.
Among the tested IQA methods, PieAPP and DISTS are trained with other datasets.
Although PieAPP achieves the best SRCC performance on both the PIPAL dataset and the TID2013, their training data are not publicly available.
Direct comparison with these methods is unfair.
LPIPS-VGG and LPIPS-Alex are trained using BAPPS, which are the same as ours; thus they are suitable for comparing the proposed algorithms.
As can be seen, the proposed SWDN obtains comparable performance on both the PIPAL dataset and TID2013 dataset.
When it comes to GAN-based distortion, the proposed SWDN achieves better performance compared with the existing IQA methods.
Especially, the SWDN outperforms the LPIPS-Alex and LPIPS-VGG.
Note that the LPIPS-Alex uses the same feature extraction backbone architecture with SWDN, which indicates that the proposed strategies are effective.

The proposed PIPAL dataset can also be used to train the SWDN network.
We split the PIPAL dataset into training and testing sets.
For the reference images, we split 100 reference images to the testing set and 150 images to the training set.
For the distorted images, we randomly split half of the distorted images of the SR distortion sub-type into the testing set, including the outputs of traditional, PSNR-oriented, and GAN-based SR algorithms.
For comparison, we train the SWDN and LPIPS-Alex using the PIPAL training dataset.
The results are shown in the last two rows of \tablename~\ref{tab:iqa_net} and are marked with ``*''.
As one can observe, compared with the versions trained using the BAPPS dataset, both the LPIPS-Alex and SWDN trained on the PIPAL dataset achieves better performance on TID2013 and the PIPAL test set, which indicates that building a large-scale dataset with GAN-based distortion can directly help to get better IQA methods.
When using the same training dataset, the proposed SWDN also outperforms LPIPS-Alex.

To investigate the behavior of SWDN as a proposal method, we conduct ablation studies to show the effect of different components.
For the baseline architecture, we replace the $l_2$ pooling layers with the original max-pooling layers and remove SWD layers.
The results are shown in \tablename~\ref{tab:SWD_ablation} and there are 8 experiments.
First, as shown by the experiments \tablename~\ref{tab:SWD_ablation}~(1,2), the $l_2$ pooling layers can improve the performance on GAN-based distortion at a small cost.
Second, we show the results with only SWD layers and the hyper-parameter $d$ varies from 1 to 5.
As can be observed, for all $d$s, the proposed SWD layer can improve the performance of GAN-based distortion.
When $d$ is set to be 3, the network with SWD layers achieves the best performance.
Thus, we choose $d=3$ for a better performance-speed trade-off.
At last, we combine the $l_2$ pooling layers and SWD layers to form the final SWDN network, and the results are shown in \tablename~\ref{tab:SWD_ablation}~(8).
The final SWDN network achieves the best performance of the GAN-based distortion type.

\section{Conclusion}
In this paper, we contribute a novel large-scale IQA dataset, namely PIPAL dataset.
PIPAL contains 250 reference images, 40 distortion types, 29k distortion images, and more than one million human ratings.
Especially, we include 24 GAN-based algorithms’ outputs as a new GAN-based distortion type.
We employ the Elo rating system to assign the Mean Opinion Scores (MOS).
Based on the PIPAL dataset, we establish benchmarks for both IQA methods and SR algorithms.
Our results indicate that existing IQA methods face challenges in evaluating perceptual IR algorithms, especially GAN-based algorithms.
With the development of IR technology, new IQA methods need to be proposed accordingly.
We also propose a new IQA network called Space Warping Difference Network, which consists of $l_2$ pooling layers and novel Space Warping Difference layers.
Experiments demonstrate the effectiveness of the proposed network.

\begin{acknowledgements}
This work is partially supported by the National Natural Science Foundation of China (61906184), Science and Technology Service Network Initiative of Chinese Academy of Sciences (KFJ-STS-QYZX-092), Shenzhen Basic Research Program (JSGG20180507182100698, CXB20110422-0032A), the Joint Lab of CAS-HK, Shenzhen Institute of Artificial Intelligence and Robotics for Society, and SenseTime Research.
\end{acknowledgements}

\clearpage
\bibliographystyle{spbasic}      
\bibliography{pipalbib.bib}   

\appendix

\section*{Appendix}

\begin{figure*}[b]
\centering
\includegraphics[width=1.0\linewidth]{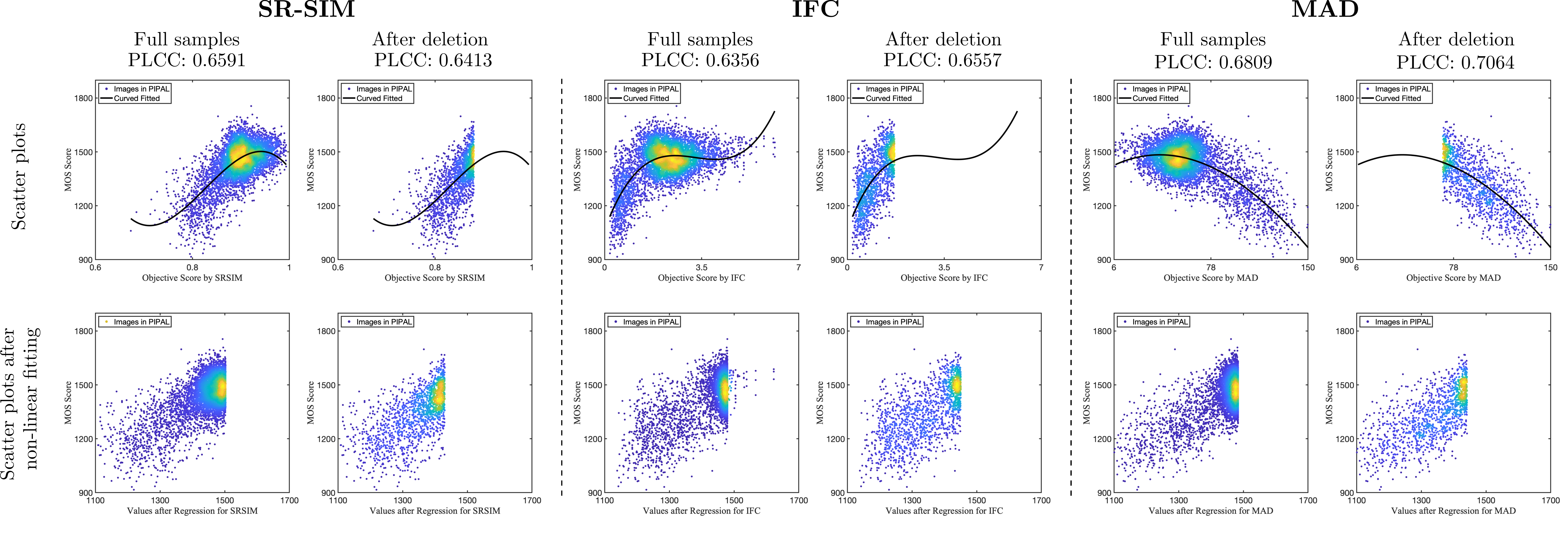}
\caption{The first line shows the scatter plots for SR-SIM, IFC and MAD on GAN-based distortion. The scatter plots after non-linear fitting are shown in the second line. For each method, the figures on the left include full samples and the figures on the right show only samples with low perceptual qualities. As can be seen, deleting the samples that are difficult for these IQA methods will not change the PLCC performance significantly.
This experiment shows that the PLCC index may overestimate the performance when the IQA method cannot effectively indicate image similarity.}
\label{fig:apd:plcc}
\end{figure*}

\section{More Results}
\label{apd:results}
In this section, we provide more benchmark results.
We evaluate IQA methods using Spearman rank order correlation coefficients (SRCC) \citep{sheikh2006statistical} and Kendall rank order correlation coefficients (KRCC) \citep{kendall1977advanced}.
These two indexes evaluate the monotonicity of the methods: whether the scores of high-quality images are higher (or lower) than low-quality images.
We also provide the Pearson linear correlation coefficient (PLCC) results.
The PLCC index evaluates the accuracy of the methods.
Before calculating the PLCC index, we perform a nonlinear regression to fit the subjective scores and the objective scores using third-order polynomials fitting.
The KRCC results are shown in \tablename~\ref{tab:apd:krcc} and the PLCC results are shown in \tablename~\ref{tab:apd:plcc}.
In the paper, we prefer to analyze SRCC and KRCC because the PLCC index may overestimate the performance when the IQA method cannot effectively indicate image similarity.
As shown in \tablename~\ref{tab:apd:plcc}, some IQA methods obtain high PLCC performance such as SR-SIM, IFC and MAD, which is inconsistent with the conclusion reached by observing SRCC performance.
We argue that this is because these methods fail to predict image quality when the Elo score is high, and then the samples with low Elo scores dominate the PLCC performance.
We show the scatter plots of SR-SIM, IFC and MAD in \figurename~\ref{fig:apd:plcc}.
As one can see, all the fitted curves of those IQA methods tend to horizontal in the area with better image quality, which means they fail to predict image qualities.
After non-linear fitting, the abscissa values of these samples are concentrated in a small interval, and thus cannot have enough influence on the calculation of the PLCC index.
We verify this phenomenon through an experiment.
In \figurename~\ref{fig:apd:plcc}, we show the situation of removing these samples and their PLCC performance has not changed significantly.
This proves that it is inappropriate to use PLCC for evaluation.
At last, we show more results for the SR benchmark in \tablename~\ref{tab:sr_evaluation}, including more algorithms and IQA methods.

\section{Details of the Distortion types}
\label{apd:distortion_details}
Recall that we have 40 distortion types and 116 levels of distortion for each reference image in the PIPAL dataset.
In addition to the traditional distortion types, we also include the outputs of a variety of real IR algorithms.
In \tablename~\ref{tab:distortions}, we present the details of these distortion types, including the selected algorithms and the parameters to create distorted images (such as the SR factors and noise levels) and the implementation details of the traditional distortions in the proposed dataset.

For the spatial warping distortion, we apply local spatial shift to each pixel on several randomly selected regions in the image.
Given a line with starting point $\vec{c}$ and termination point $\vec{m}$, the per-pixel shift map around this line can be computed using the following formula:
$$
\vec{u}=\vec{x}-\bigg(\frac{\vec{r}^2-|\vec{x}-\vec{c}|^2}{(\vec{r}^2-|\vec{x}-\vec{c}|^2)+|\vec{m}-\vec{c}|^2)}\bigg)^2(\vec{m}-\vec{c}),
$$
where $\vec{u}$ is the original location of the point $\vec{x}$, and $\vec{r}$ is the radius of influence.
The source pixels are then resampled to form the target pixel.
For an input image, we randomly select $n$ warping points and warping distance $s$ (distance between point $\vec{c}$ and $\vec{m}$) to perform locally spatial warping.
We set four distortion levels for spatial warping with parameters $n=\{4,16,32,64\}$, $s=\{2,3,4,6\}$ and $\vec{r}=\{15,25,35,60\}$.

\begin{figure}[t]
    \centering
    \includegraphics[width=1.0\linewidth]{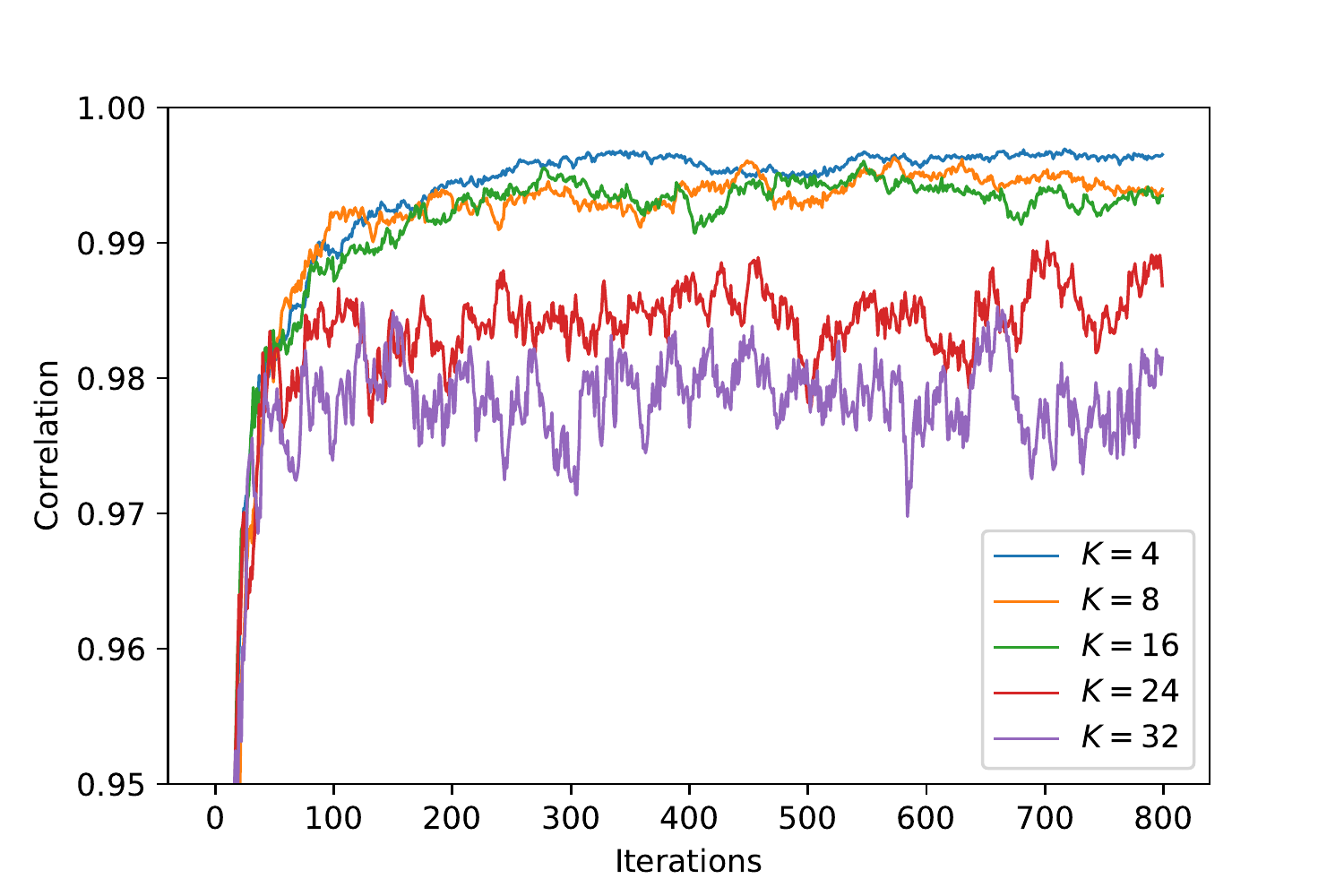}
    \caption{The convergence curve of the Elo rating system with different $K$. As can be seen, experiments using larger $K$ converge faster. However, experiments with smaller $K$ converge more stable.}\label{fig:Elo_converge}
\end{figure}  

\begin{figure}[t]
    \centering
    \includegraphics[width=1.0\linewidth]{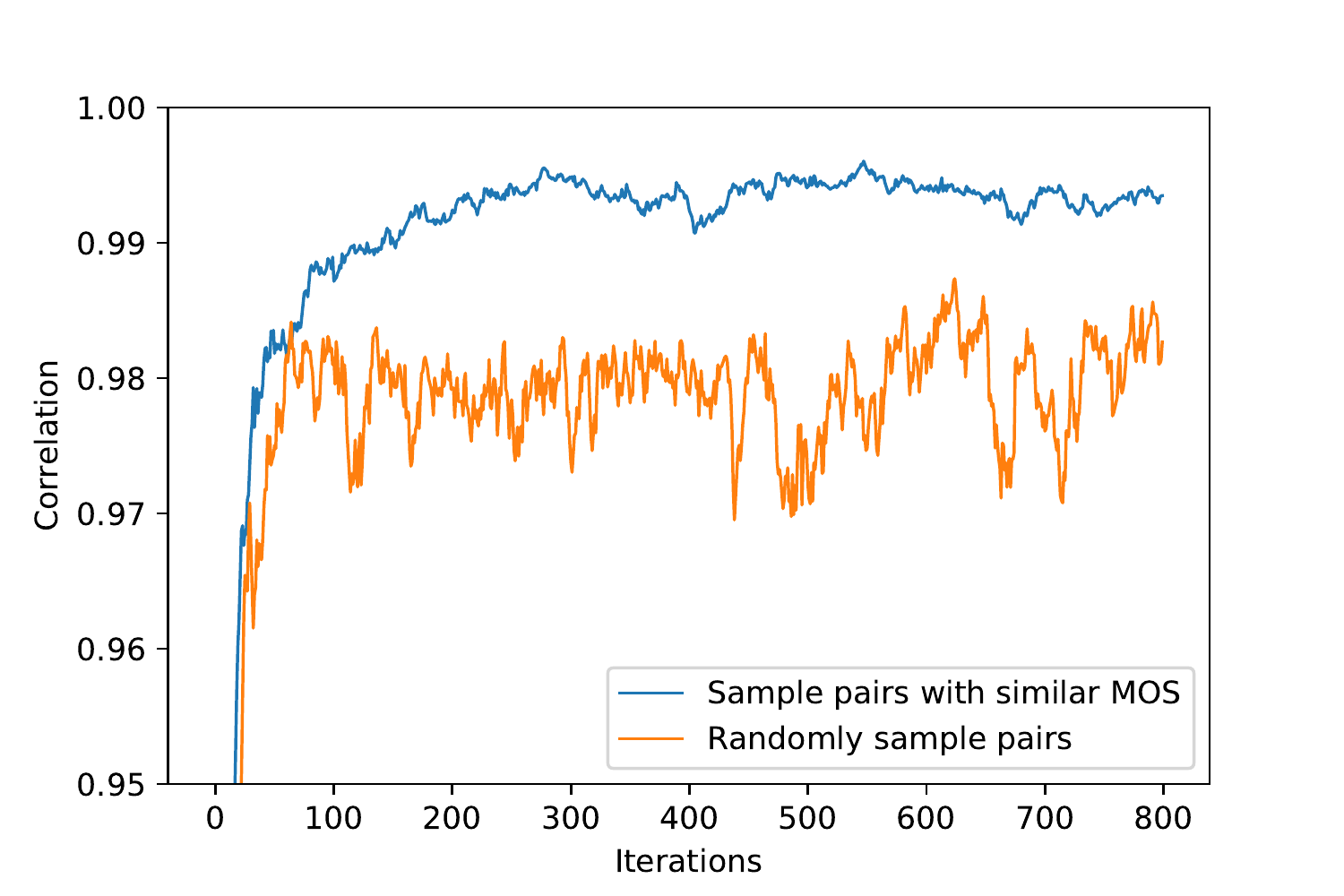}
    \caption{The convergence curve of the Elo rating system with different $K$. As can be seen, experiments using larger $K$ converge faster. However, experiments with smaller $K$ converge more stable.}\label{fig:Elo_choice}
\end{figure}  

\begin{figure}[t]
    \centering
    \includegraphics[width=1.0\linewidth]{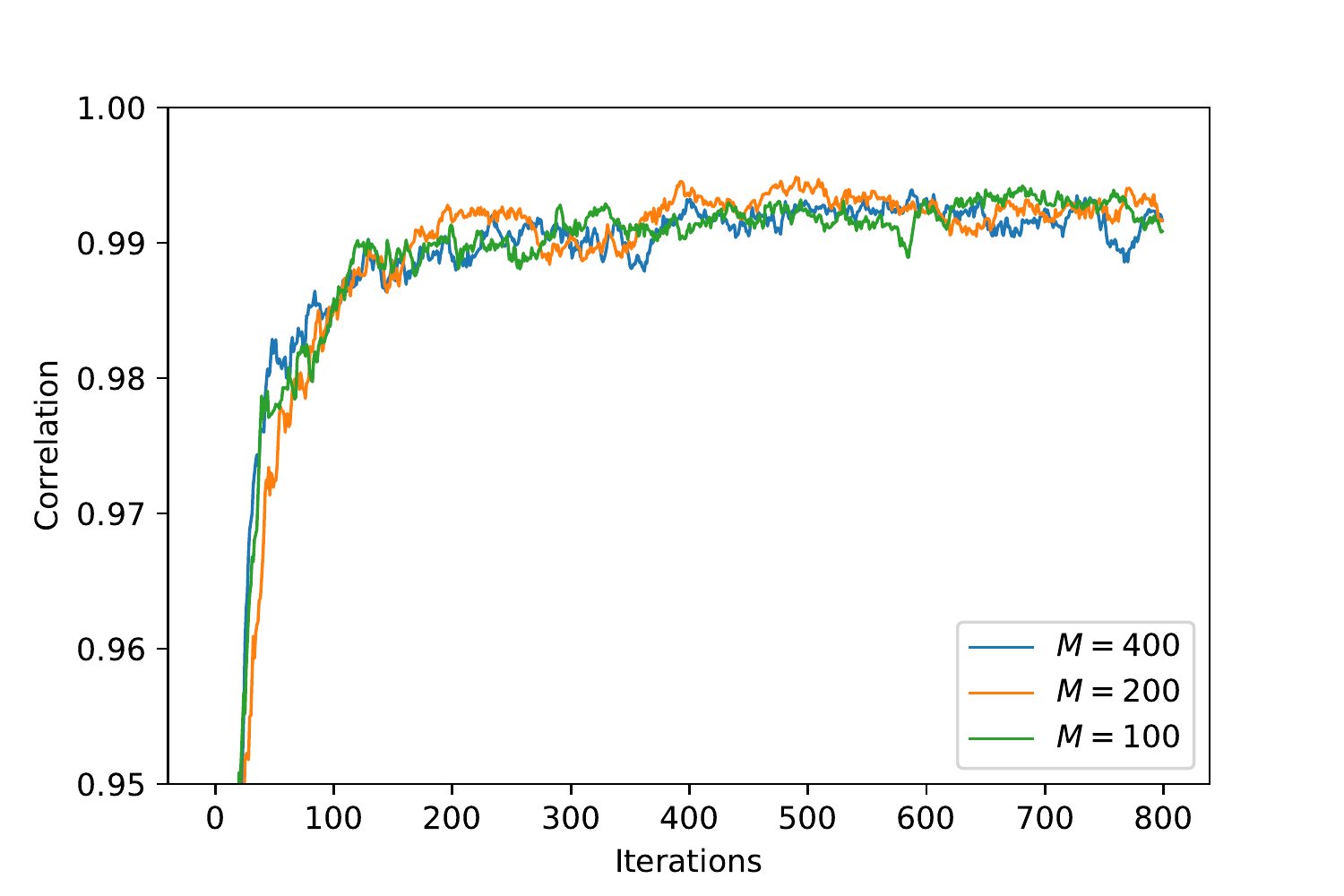}
    \caption{The convergence curve of the Elo rating system with different $M$. As one can see, the value of $M$ does not affect the ranking accuracy.}\label{fig:Elo_base}
\end{figure}  

\begin{figure}[t]
    \centering
    \includegraphics[width=1.0\linewidth]{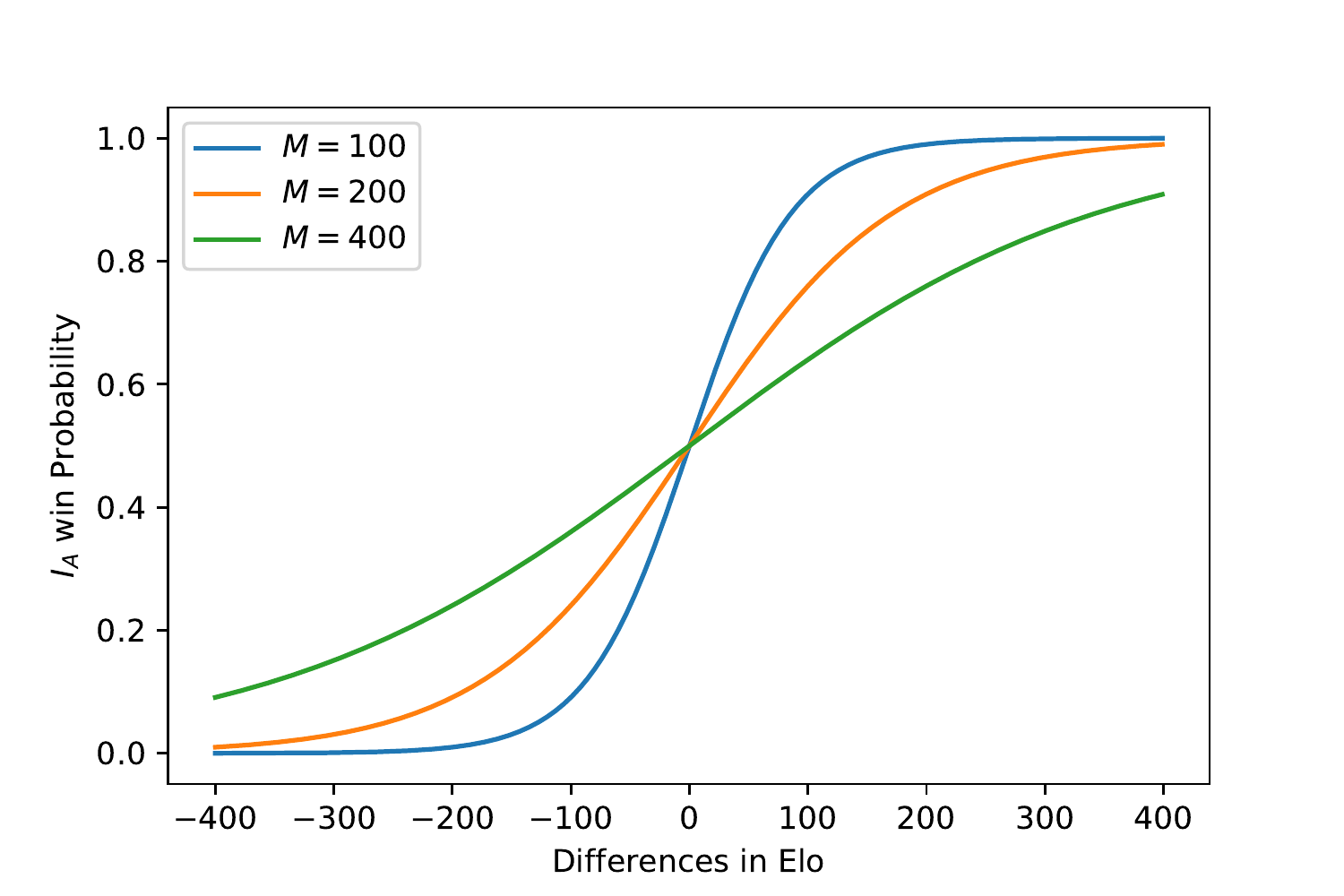}
    \caption{The probability estimation with different $M$. Tuning $M$ only affects the probability curve’s kurtosis.}\label{fig:Elo_m}
\end{figure}  

\begin{figure}[t]
    \centering
    \includegraphics[width=1.0\linewidth]{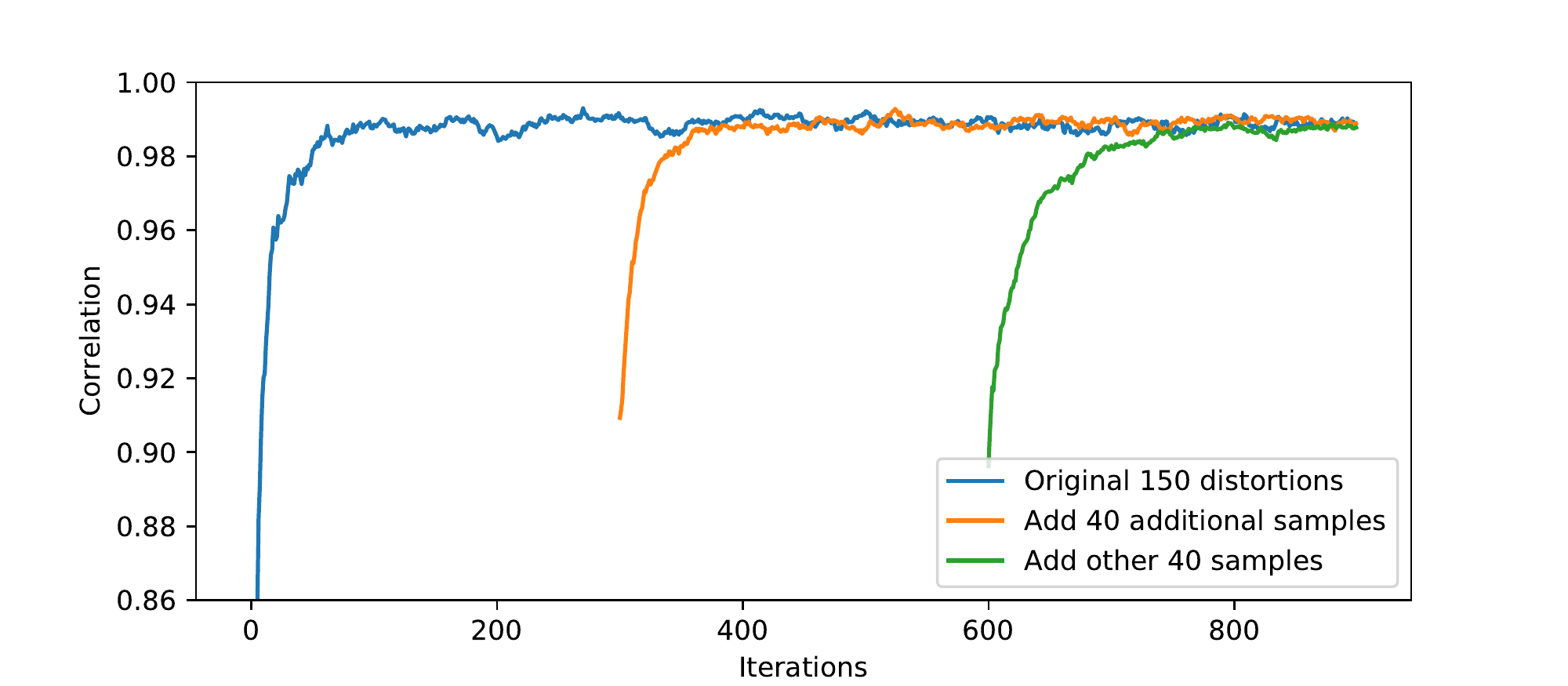}
    \caption{The convergence curve of the Elo rating system with different sampling strategies. As can be seen, sampling pairs with similar MOS scores has better convergence performance.}\label{fig:Elo_extend}
\end{figure}  

\section{Reliability and Expandability of Elo System}
\label{sec:Elo}
In the PIPAL dataset, we employ the Elo rating system to obtain mean opinion scores (MOS) for distorted images.
The Elo rating system is a statistic-based rating method and is able to bring pairwise preference probability and rating system together.
In this section, we show the effectiveness and expandability of the Elo rating system with quantitative experiments.
We also performed ablation studies on some of the hyper-parameters of the Elo system.

To validate the Elo rating system, we build a simulation set consisting of 150 populations and assign the ground truth scores for them.
These scores follow the constraints that:
(1) The ranking is transferable. if $A>B$ and $B>C$, then $A>C$,
(2) Given a ranking, then the expectations of any comparison should be the same as the ranking order. If $A>B>C$, then the probability of $A>C$ is higher than the probability of $A>B$.
We simulate the process of scoring these populations using the Elo system.
We then use the Spearman rank correlation coefficient between the Elo rating results and the ground truth to measure the effectiveness of the Elo system.
We first evaluate the Elo system with different parameters of $K$.
As can be seen in \figurename~\ref{fig:Elo_converge}, all the experiments converge quickly.
When $K$ becomes smaller, the convergence speed becomes slower but more stable.
Considering the trade-off between convergence speed and the rating accuracy, we use $K = 16$ to build our dataset.
In \figurename~\ref{fig:Elo_choice}, we show the experiments with different sample strategies.
Recall that in our dataset, we intentionally select the distorted images with similar Elo scores for users to make judgements.
The experiments show that selecting images with similar scores converge quicker and better than random selecting.
We also conduct experiments to show the effect of parameter $M$.
As can be seen in \figurename~\ref{fig:Elo_base}, the choice of $M$ does not affect the convergence performance and accuracy of the Elo system.
However, changing $M$ enables us to control win probability estimation based on Elo scores.
\figurename~\ref{fig:Elo_m} illustrates how tuning $M$ affects probability estimation by changing the curve’s kurtosis.
In general, a lower $M$ indicates that we can be more confident in our predictions since they’re less susceptible to random events.
The same conclusion can be found in some prior works \cite{elom}.
At last, we experimentally show the expandability of the Elo system.
We first build a set with populations of 150 and perform Elo rating.
When the Elo rating process converges, we add the other 40 populations and keep updating the Elo scores for all of the samples.
The curve is shown in \figurename~\ref{fig:Elo_extend}, as one can see the Elo system quickly adjusts according to the new samples without losing old Elo scores.
This experiment demonstrates the used Elo system is expandable.

\begin{table*}[t]
    \centering
    \footnotesize
    \resizebox{1.0\linewidth}{!}{
    \begin{tabular}{
    p{2.0cm}
    p{2.1cm}<{\centering}
    p{2.1cm}<{\centering}
    p{2.1cm}<{\centering}
    p{2.1cm}<{\centering}
    p{2.1cm}<{\centering}
    p{2.1cm}<{\centering}}
    \toprule
        \multirow{2}{*}{\textbf{Method}} &
        \textbf{Traditional} &
        \multirow{2}{*}{\textbf{Denoising}} &
        \multirow{2}{*}{\textbf{SR Full}} &
        \textbf{Traditional} &
        \textbf{PSNR.} &
        \textbf{\textit{GAN-based}}
        \\
         &
        \textbf{Distortion} &
         &
         &
        \textbf{SR} &
        \textbf{SR} &
        \textbf{\textit{SR}}
        \\
        \hline
        PSNR	$\uparrow$	                &	0.2427 	&	0.3072 	&	0.2810 	&	0.3245 	&	0.3754 	&	0.1939 	\\
        NQM	$\uparrow$	                    &	0.1764 	&	0.3948 	&	0.3314 	&	0.3688 	&	0.4582 	&	0.2340 	\\
        UQI	$\uparrow$	                    &	0.2361 	&	0.4399 	&	0.3689 	&	0.4222 	&	0.5095 	&	0.2291 	\\
        SSIM	$\uparrow$	                &	0.2664 	&	0.4763 	&	0.3640 	&	0.4029 	&	0.4931 	&	0.2315 	\\
        MS-SSIM	$\uparrow$	                &	0.2743 	&	0.4988 	&	0.3962 	&	0.4558 	&	0.5498 	&	0.2631 	\\
        IFC	$\uparrow$  	                &	0.2576 	&	\textbf{0.5405} 	&	0.4060 	&	\textbf{0.5024} 	&	\textbf{0.6269} 	&	0.2166 	\\
        VIF	$\uparrow$	                    &	0.3166 	&	\textbf{0.5273} 	&	0.4243 	&	\textbf{0.4902} 	&	\textbf{0.5849} 	&	0.2664 	\\
        VSNR-FR	$\uparrow$	                &	0.2811 	&	0.4144 	&	0.3572 	&	0.4259 	&	0.5054 	&	0.2146 	\\
        RFSIM	$\uparrow$	                &	0.2325 	&	0.3066 	&	0.2909 	&	0.3114 	&	0.3815 	&	0.2028 	\\
        GSM	$\uparrow$	                    &	0.3946 	&	0.4239 	&	0.3736 	&	0.4199 	&	0.4905 	&	0.2395 	\\
        SR-SIM	$\uparrow$	                &	\textbf{0.4287} 	&	0.4797 	&	\textbf{0.4336} 	&	0.4600 	&	0.5470 	&	0.3211 	\\
        FSIM	$\uparrow$	                &	0.4060 	&	0.4928 	&	0.4167 	&	0.4563 	&	0.5363 	&	0.2806 	\\
        FSIM$_\mathcal{C}$	$\uparrow$	    &	0.4023 	&	0.4913 	&	0.4149 	&	0.4558 	&	0.5354 	&	0.2784 	\\
        VSI	$\uparrow$	                    &	0.3435 	&	0.3990 	&	0.3818 	&	0.4202 	&	0.4943 	&	0.2527 	\\
        MAD	$\downarrow$	                &	0.2615 	&	0.5040 	&	0.3866 	&	0.4744 	&	0.5593 	&	0.2387 	\\
        LPIPS-Alex	$\downarrow$	        &	0.2803 	&	0.5216 	&	0.4325 	&	0.4199 	&	0.5302 	&	\textbf{0.3330} 	\\
        LPIPS-VGG	$\downarrow$    	    &	0.4191 	&	0.4747 	&	0.3873 	&	0.3740 	&	0.4804 	&	0.3323 	\\
        PieAPP	$\downarrow$	            &	\textbf{0.5013} 	&	\textbf{0.5455} 	&	\textbf{0.5252} 	&	\textbf{0.5368} 	&	\textbf{0.6106} 	&	\textbf{0.3865} 	\\
        DISTS	$\downarrow$	            &	\textbf{0.4445} 	&	0.5193 	&	\textbf{0.4639} 	&	0.4712 	&	0.5705 	&	\textbf{0.3840} 	\\
        \underline{NIQE}	$\downarrow$	&	0.0741 	&	0.0040 	&	0.0211 	&	0.0401 	&	0.1014 	&	0.0102 	\\
        \underline{Ma}	$\uparrow$	        &	0.3170 	&	0.3375 	&	0.2603 	&	0.4301 	&	0.5195 	&	0.0363 	\\
        \underline{PI}	$\downarrow$	    &	0.2498 	&	0.2111 	&	0.1368 	&	0.3316 	&	0.4020 	&	0.0126 	\\
    \toprule
    \end{tabular}}
    \caption{The KRCC results with respect to different distortion sub-types.
    $\uparrow$ means the higher the better while $\downarrow$ means the lower the better. Higher coefficient matches perceptual score better. We indicate the top 3 performance with \textbf{blod} values.
    }
    \label{tab:apd:krcc}
\end{table*}  

\begin{table*}[t]
    \centering
    \footnotesize
    \resizebox{1.0\linewidth}{!}{
    \begin{tabular}{
    p{2.0cm}
    p{2.1cm}<{\centering}
    p{2.1cm}<{\centering}
    p{2.1cm}<{\centering}
    p{2.1cm}<{\centering}
    p{2.1cm}<{\centering}
    p{2.1cm}<{\centering}}
    \toprule
        \multirow{2}{*}{\textbf{Method}} &
        \textbf{Traditional} &
        \multirow{2}{*}{\textbf{Denoising}} &
        \multirow{2}{*}{\textbf{SR Full}} &
        \textbf{Traditional} &
        \textbf{PSNR.} &
        \textbf{\textit{GAN-based}}
        \\
         &
        \textbf{Distortion} &
         &
         &
        \textbf{SR} &
        \textbf{SR} &
        \textbf{\textit{SR}}
        \\
        \hline
            PSNR	$\uparrow$	                &	0.3639 	&	0.5055 	&	0.4359 	&	0.4705 	&	0.5625 	&	0.3931 	\\
            NQM	$\uparrow$                  	&	0.2868 	&	0.6059 	&	0.5129 	&	0.5279 	&	0.6743 	&	0.5002 	\\
            UQI	$\uparrow$	                    &	0.3619 	&	0.6666 	&	0.5854 	&	0.5991 	&	0.7367 	&	0.5755 	\\
            SSIM	$\uparrow$	                &	0.4013 	&	0.6986 	&	0.5613 	&	0.5765 	&	0.7138 	&	0.5126 	\\
            MS-SSIM	$\uparrow$              	&	0.4084 	&	0.7361 	&	0.6034 	&	0.6532 	&	0.7779 	&	0.6022 	\\
            IFC	$\uparrow$                  	&	0.3888 	&	0.7549 	&	0.6293 	&	\textbf{0.7044} 	&	\textbf{0.8582} 	&	\textbf{0.6356} 	\\
            VIF	$\uparrow$	                    &	0.4575 	&	\textbf{0.7603} 	&	0.6342 	&	\textbf{0.6932} 	&	\textbf{0.8282} 	&	0.6069 	\\
            VSNR-FR	$\uparrow$	                &	0.4165 	&	0.6236 	&	0.5298 	&	0.6072 	&	0.7163 	&	0.4495 	\\
            RFSIM	$\uparrow$              	&	0.3508 	&	0.4765 	&	0.4394 	&	0.4520 	&	0.5670 	&	0.3761 	\\
            GSM	$\uparrow$	                    &	0.5762 	&	0.6372 	&	0.5502 	&	0.6015 	&	0.7021 	&	0.4612 	\\
            SR-SIM	$\uparrow$	                &	\textbf{0.6273} 	&	0.7165 	&	\textbf{0.6681} 	&	0.6531 	&	0.7915 	&	\textbf{0.6591} 	\\
            FSIM	$\uparrow$              	&	0.5994 	&	0.7252 	&	0.6279 	&	0.6436 	&	0.7668 	&	0.5816 	\\
            FSIM$_\mathcal{C}$	$\uparrow$  	&	0.5983 	&	0.7237 	&	0.6260 	&	0.6431 	&	0.7660 	&	0.5793 	\\
            VSI	$\uparrow$                  	&	0.5231 	&	0.6113 	&	0.5735 	&	0.6018 	&	0.7159 	&	0.4998 	\\
            MAD	$\downarrow$                	&	0.3987 	&	\textbf{0.7605} 	&	0.6367 	&	0.6737 	&	0.8096 	&	\textbf{0.6809} 	\\
            LPIPS-Alex	$\downarrow$        	&	0.4245 	&	0.7436 	&	0.6376 	&	0.6017 	&	0.7491 	&	0.6101 	\\
            LPIPS-VGG	$\downarrow$        	&	0.5960 	&	0.6793 	&	0.5458 	&	0.5424 	&	0.6821 	&	0.5059 	\\
            PieAPP	$\downarrow$            	&	\textbf{0.6996} 	&	\textbf{0.7765} 	&	\textbf{0.6982} 	&	\textbf{0.7311} 	&	\textbf{0.8336} 	&	0.5626 	\\
            DISTS	$\downarrow$            	&	\textbf{0.6402} 	&	0.7412 	&	\textbf{0.6467} 	&	0.6765 	&	0.8046 	&	0.5988 	\\
            \underline{NIQE}	$\downarrow$	&	0.1506 	&	0.1029 	&	0.0785 	&	0.1342 	&	0.1322 	&	0.0650 	\\
            \underline{Ma}	$\uparrow$	        &	0.4876 	&	0.5027 	&	0.3645 	&	0.6312 	&	0.7067 	&	0.0201 	\\
            \underline{PI}	$\downarrow$    	&	0.3572 	&	0.3487 	&	0.1760 	&	0.4605 	&	0.5565 	&	0.0420 	\\
    \toprule
    \end{tabular}}
    \caption{The PLCC results with respect to different distortion sub-types.
    $\uparrow$ means the higher the better while  $\downarrow$ means the lower the better. Higher coefficient matches perceptual score better. We indicate the top 3 performance with \textbf{blod} values.
    }
    \label{tab:apd:plcc}
\end{table*}  

\begin{table*}
\begin{center}
    \footnotesize
    {\resizebox{\linewidth}{!}{
  \begin{tabular}{
  p{2.6cm}
  p{1.0cm}
  p{1.2cm}
  p{1.2cm}
  p{1.2cm}
  p{1.2cm}
  p{1.2cm}
  p{1.2cm}
  p{1.2cm}
  p{1.4cm}
  p{1.4cm}
  p{1.4cm}
  }
        \toprule
        Method &
        Year &
        PSNR $\uparrow$ &
        ~SSIM $\uparrow$ &
        ~IFC $\uparrow$ &
        FSIM $\uparrow$ &
        ~\underline{Ma} $\uparrow$ &
        \underline{NIQE} $\downarrow$ &
        ~~\underline{PI} $\downarrow$ &
        LPIPS $\downarrow$ &
        PieAPP $\downarrow$ &
        ~~MOS $\uparrow$ \\
        \specialrule{0em}{1pt}{1pt}
        \hline
        \specialrule{0em}{1pt}{1pt}
YY	&	2013	&	23.35 	&	0.6897 	&	2.1900 	&	0.8018 	&	4.5487 	&	6.4174 	&	5.9344 	&	0.3574 	&	2.4647 	&	1367.71 	\\
TSG	&	2013	&	23.55 	&	0.6775 	&	2.4256 	&	0.7812 	&	4.1298 	&	6.4164 	&	6.1433 	&	0.3570 	&	2.8931 	&	1387.24 	\\
A+	&	2014	&	23.82 	&	0.6919 	&	2.5547 	&	0.7908 	&	4.3852 	&	6.3645 	&	5.9897 	&	0.3491 	&	2.4406 	&	1354.52 	\\
SRCNN	&	2014	&	23.93 	&	0.6966 	&	2.4564 	&	0.7991 	&	4.6094 	&	6.5657 	&	5.9781 	&	0.3316 	&	2.1338 	&	1363.68 	\\
FSRCNN	&	2016	&	24.07 	&	0.7013 	&	2.4428 	&	0.8018 	&	4.6686 	&	6.9985 	&	6.1649 	&	0.3281 	&	2.0569 	&	1367.49 	\\
VDSR	&	2016	&	24.13 	&	0.6984 	&	2.5377 	&	0.7955 	&	4.7799 	&	7.4436 	&	6.3319 	&	0.3484 	&	2.1035 	&	1364.90 	\\
EDSR	&	2017	&	25.17 	&	0.7541 	&	3.1431 	&	0.8362 	&	5.7634 	&	6.4560 	&	5.3463 	&	0.3016 	&	1.5206 	&	1447.44 	\\
EnhanceNet	&	2017	&	22.20 	&	0.6245 	&	2.0267 	&	0.8123 	&	\textbf{8.5480} 	&	4.6177 	&	3.0348 	&	0.3381 	&	1.3421 	&	1434.51 	\\
RCAN	&	2018	&	\textbf{25.21} 	&	\textbf{0.7569} 	&	\textbf{3.1696} 	&	\textbf{0.8381} 	&	5.9260 	&	6.4121 	&	5.2430 	&	0.2992 	&	1.4662 	&	1455.31 	\\
SFTGAN	&	2018	&	22.85 	&	0.6576 	&	2.1023 	&	0.8189 	&	8.3307 	&	4.6013 	&	3.1353 	&	0.3118 	&	1.3007 	&	1437.38 	\\
BOE R1	&	2018	&	24.30 	&	0.7181 	&	2.8459 	&	0.8260 	&	6.2004 	&	5.2071 	&	4.5033 	&	0.2960 	&	1.6064 	&	1419.83 	\\
BOE R2	&	2018	&	23.67 	&	0.6955 	&	2.6801 	&	0.8252 	&	7.8473 	&	4.1193 	&	3.1360 	&	0.2950 	&	1.2972 	&	1452.88 	\\
BOE R3	&	2018	&	22.68 	&	0.6582 	&	2.5137 	&	0.8226 	&	8.5209 	&	\textbf{3.7945} 	&	2.6368 	&	0.2933 	&	0.9653 	&	1481.51 	\\
EPSR R1	&	2018	&	24.65 	&	0.7206 	&	2.7449 	&	0.8326 	&	7.1709 	&	4.4977 	&	3.6634 	&	\textbf{0.2513} 	&	1.4958 	&	1446.84 	\\
EPSR R2	&	2018	&	24.10 	&	0.6985 	&	2.5294 	&	0.8350 	&	8.0417 	&	4.0555 	&	3.0069 	&	\textbf{0.2542} 	&	1.3162 	&	1463.99 	\\
EPSR R3	&	2018	&	22.49 	&	0.6496 	&	2.0185 	&	0.8224 	&	8.4434 	&	3.8855 	&	2.7211 	&	0.2918 	&	0.9854 	&	1494.39 	\\
PESR	&	2018	&	22.38 	&	0.6676 	&	2.1698 	&	0.8296 	&	8.4818 	&	4.1061 	&	2.8122 	&	0.2654 	&	\textbf{0.9344} 	&	\textbf{1517.32} 	\\
ESRGAN $\alpha=0.4$	&	2018	&	\textbf{25.45} 	&	\textbf{0.7646} 	&	\textbf{3.2376} 	&	\textbf{0.8427} 	&	5.8255 	&	6.4889 	&	5.3317 	&	0.2957 	&	1.4676 	&	1453.47 	\\
ESRGAN $\alpha=0.8$	&	2018	&	\textbf{25.45} 	&	\textbf{0.7646} 	&	\textbf{3.2376} 	&	\textbf{0.8427} 	&	5.8256 	&	6.4888 	&	5.3316 	&	0.2957 	&	1.4676 	&	1450.42 	\\
ESRGAN $\alpha=1.0$	&	2018	&	22.51 	&	0.6566 	&	2.2872 	&	0.8300 	&	8.3424 	&	4.7821 	&	3.2198 	&	\textbf{0.2517} 	&	\textbf{0.8641} 	&	\textbf{1534.25} 	\\
RankSRGAN (a)	&	2019	&	22.11 	&	0.6392 	&	2.0309 	&	0.8168 	&	\textbf{8.6882} 	&	3.8155 	&	\textbf{2.5636} 	&	0.2755 	&	\textbf{0.9017} 	&	\textbf{1518.29} 	\\
RankSRGAN (b)	&	2019	&	22.55 	&	0.6488 	&	2.1198 	&	0.8226 	&	\textbf{8.5341} 	&	\textbf{3.7635} 	&	\textbf{2.6147} 	&	0.2697 	&	0.9668 	&	1503.52 	\\
RankSRGAN (c)	&	2019	&	22.62 	&	0.6485 	&	2.1245 	&	0.8230 	&	8.5260 	&	\textbf{3.7169} 	&	\textbf{2.5955} 	&	0.2705 	&	0.9718 	&	1510.56 	\\
        \toprule
    \end{tabular}
    }}
    \caption{The $\times4$ SR results. The years of publication are also provided. $\uparrow$ means the higher the better while  $\downarrow$ means the lower the better. The \textbf{bolded} values are the top 3 values.
    }
    \label{tab:sr_evaluation}
\end{center}
\end{table*}  

\begin{table*}[]
    \centering
{\resizebox{\linewidth}{!}{
\begin{tabular}{
    p{3.3cm}
    p{5.0cm}
    p{8cm}
}
        \textbf{Distortion Sub-type} & \textbf{Distortion types} & \textbf{Implementation Detail} \\
        \hline
        
        \multirow{15}{*}{Traditional Distortions}
        & 1. Median filter denoising & The noise levels of the salt pepper noise are 0.08 and 0.16. \\
        & 2. Linear motion blur & fixed width with two different directions. \\
        & 3. JPEG and JPEG 2000 & The quality levels for JPEG: 10 and 20. The quality levels for JPEG 2000 are 20, 40 and 60. \\
        & 4. Color quantization & We first use the MATLAB \texttt{multithresh} function to segment the image and then quantizing with the intensity levels of 3 and 7.\\
        & 5. Gaussian noise & The Gaussian noise levels are 10, 15 and 25.\\
        & 6. Gaussian blur & The Gaussian blur with $\sigma=1$ $\sigma=1.8$ and $\sigma=3.2$.\\
        & 7. Bilateral filtering & The range parameter and the spatial parameter are set to $\{0.3,0.35\}$ and $\{1.7, 3.0\}$. \\
        & 8. Spatial warping & See details in Appendix Sec.~\ref{apd:distortion_details}. \\
        & 9. Comfort noise & We use the implementation from TID2013 and we have 4 levels. \\
        \hline
        \multirow{5}{3.2cm}{Traditional SR}
         & 10. Interpolation & Bicubic upsampling $\times2$ and $\times3$.  \\
         & 11. A+ \citep{a+2014} & SR factors of $\times2$, $\times3$ and $\times4$.  \\
         & 12. YY \citep{yy2013} & SR factors of $\times2$, $\times3$ and $\times4$.  \\
         & 13. TSG \citep{tsg2013} & SR factors of $\times2$, $\times3$ and $\times4$.  \\
         & 14. YWHM \citep{ywhm2010} & SR factor of $\times2$.  \\
        \hline
        \multirow{5}{3.2cm}{PSNR-orainted SR}
         & 15. SRCNN \citep{srcnn2014} & SR factors of $\times2$, $\times3$ and $\times4$. \\
         & 16. FSRCNN \citep{fsrcnn2016} & SR factors of $\times2$, $\times3$ and $\times4$. \\
         & 17. VDSR \citep{vdsr2016} & SR factors of $\times2$, $\times3$ and $\times4$. \\
         & 18. EDSR \citep{edsr2017} & SR factors of $\times2$, $\times3$ and $\times4$. \\
         & 19. RCAN \citep{rcan2018} & SR factors of $\times2$, $\times3$, $\times4$ and $\times8$. \\
         \hline
         \multirow{5}{3.2cm}{SR with kernel mismatch}
         & \multirow{5}{4.8cm}{20. SFTMD \citep{gu2019blind}} & For $\times2$ SR, the LR images are blurred with Gaussian blur with $\sigma=1.0$, SR using the Gaussian kernel with $\sigma=0.7, 0.85,1.0,1.1,1.3$. For $\times4$ SR, the LR images are blurred with Gaussian blur with $\sigma=2.0$, SR using the Gaussian kernel with $\sigma=1.7, 1.9,2.0,2.1,2.25$.\\
         \hline
        \multirow{17}{3.2cm}{GAN-based SR}
         & 21. EnhanceNet \citep{enhancenet2017} & SR factors of $\times4$.\\
         & 22. SRGAN \citep{srgan2017} & SR factors of $\times4$, $\times6$ and $\times8$.\\
         & 23. SFTGAN \citep{sftgan2018} & SR factors of $\times4$.\\
         & 24. ESRGAN \citep{wang2018esrgan} & SR factors of $\times4$, $\times6$ and $\times8$. For $\times4$ SR, we use the network interpolation with $\alpha=0.4,0.6,0.8,1.0$ for four different GAN effects.\\
         & 25. BOE \citep{boe2018} & SR factors of $\times4$. We choose three models $R_1$, $R_2$ and $R_3$ released by the authors for different GAN effects.\\
         & 26. EPSR \citep{epsr2018} & SR factors of $\times4$. We choose three models $R_1$, $R_2$ and $R_3$ released by the authors for different GAN effects.\\
         & 27. PESR \citep{pesr2018} & SR factors of $\times4$. We use the network interpolation with $\alpha=0.5,1.0$ for two different GAN effects.\\
         & 28. EUSR \citep{eusr2018} & SR factors of $\times4$.\\
         & 29. MCML \citep{mcml2018} & SR factors of $\times4$.\\
         & 30. RankSRGAN \citep{zhang2019ranksrgan} & SR factors of $\times4$. We choose three different models (RankSRGAN-PI, RankSRGAN-MA and RankSRGAN-NIQE) for different GAN effects.\\
         \hline
         \multirow{5}{*}{Denoising}
         & 31. DnCNN \citep{dncnn} & Gaussian noise removal, the noise levels: 25 and 50.\\
         & 32. FFDNet \citep{ffdnet2018} & Gaussian noise removal, the noise levels: 25, 50 and 80.\\
         & 33. TWSC \citep{twsc} & Gaussian noise removal, the noise levels: 25, 50 and 80.\\
         & 34. BM3D \citep{bm3d} & Gaussian noise removal, the noise levels: 25, 50 and 80.\\
         & 35. ARCNN \citep{arcnn} & JPEG compression removal, the quality levels: 10 and 30.\\
        \hline
        \multirow{12}{3cm}{SR and Denoising Joint Problem}
         & 36. BM3D + EDSR & We first perform Gaussian noise removal with noise levels of 25 and 50 with BM3D, and then perform EDSR with SR factors of 2 and 3.\\
         & 37. DnCNN + EDSR & We first perform Gaussian noise removal with noise levels of 25 and 50 with DnCNN, and then perform EDSR with SR factors of 2 and 3.\\
         & 38. ARCNN + EDSR & We first perform JPEG compression removal with quality levels of 10, 20 and 40 with ARCNN, and then perform EDSR with SR factor of 2.\\
         & 39. noise + EDSR & We first add Gaussian noise with noise levels: 1.5 and 3, and then perform SR with EDSR with SR factor of 3.\\
         & 40. noise + ESRGAN & We first add Gaussian noise with noise level 0.1, and then perform SR with ESRGAN with SR factor of 4.\\
        \toprule
    \end{tabular}
}}
    \caption{Details of the included distortion types in PIPAL dataset.}
    \label{tab:distortions}
\end{table*}  

\end{document}